\documentclass[12pt,a4paper,fleqn]{article}
\usepackage{tikz}
\usepackage{caption}
\usepackage{lscape}

\usepackage{fullpage}
\usepackage{amsmath}
\usepackage{amsthm}
\usepackage{amssymb}
\usepackage{amscd}
\usepackage{plain}
\usepackage{graphicx}
\usepackage{t1enc,epsfig}
\usepackage[mathscr]{eucal}
\usepackage{epstopdf}
\usepackage[german,english]{babel}
\usepackage{stmaryrd}

\usepackage{xcolor}
\usepackage{float}

\usepackage[colorlinks, allcolors=blue]{hyperref}
\usepackage[normalem]{ulem}

\renewcommand{\thefootnote}{}

\numberwithin{equation}{section}

\begin{document}

\def\diy{\displaystyle}

\def\a{{\alpha}}
\def\b{{\beta}}
\def\Gam{{\Gamma}}
\def\gam{{\gamma}}
\def\del{{\delta}}
\def\eps{{\epsillon}}
\def\veps{{\varepsilon}}
\def\del{{\delta}}
\def\veps {\varepsilon}
\def\vphi{{\varphi}}
\def\vrho{\varrho}
\def\ups{{\upsilon}}
\def\Om {\Omega}

\def\pl {\partial}

\def\rd{{\rm d}} \def\re{{\rm e}} \def\ri{{\rm i}}
\def\rt{{\rm t}} \def\rv{{\rm v}} \def\rw{{\rm w}}
\def\rx{{\rm x}} \def\ry{{\rm y}} \def\rz{{\rm z}}
\def\rD{{\rm D}}
\def\rO{{\rm O}} \def\rP{{\rm P}} \def\rV{{\rm V}}

\def\bD{{\mathbf D}}

\def\ovphi{\ov\vphi}
\def\ophi{\ov\phi}

\def\bbA{{\mathbb A}} \def\bbB{{\mathbb B}} \def\bbC{{\mathbb C}}
\def\bbD{{\mathbb D}} \def\bbE{{\mathbb E}} \def\bbF{{\mathbb F}}
\def\bbG{{\mathbb G}} \def\bbH{{\mathbb H}}
\def\bbL{{\mathbb L}} \def\bbM{{\mathbb M}} \def\bbN{{\mathbb N}}
\def\bbP{{\mathbb P}} \def\bbQ{{\mathbb Q}}
\def\bbR{{\mathbb R}} \def\bbS{{\mathbb S}} \def\bbT{{\mathbb T}}
\def\bbU{{\mathbb U}}
\def\bbV{{\mathbb V}} \def\bbW{{\mathbb W}} \def\bbZ{{\mathbb Z}}

\def\bbZd{\bbZ^d}

\def\cA{{\mathcal A}} \def\cB{{\mathcal B}} \def\cC{{\mathcal C}}
\def\cD{{\mathcal D}} \def\cE{{\mathcal E}} \def\cF{{\mathcal F}}
\def\cG{{\mathcal G}} \def\cH{{\mathcal H}} \def\cJ{{\mathcal J}}
\def\cP{{\mathcal P}} \def\cT{{\mathcal T}} \def\cW{{\mathcal W}}
\def\cX{{\mathcal X}} \def\cY{{\mathcal Y}} \def\cZ{{\mathcal Z}}

\def\tA{{\tt A}} \def\tB{{\tt B}}
\def\tC{{\tt C}} \def\tF{{\tt F}} \def\tH{{\tt H}}
\def\tO{{\tt O}}
\def\tP{{\tt P}} \def\tQ{{\tt Q}} \def\tR{{\tt R}}
\def\tS{{\tt S}} \def\tT{{\tt T}}

\def\tx{{\tt x}} \def\ty{{\tt y}}

\def\t0{{\tt 0}} \def\t1{{\tt 1}}

\def\be{{\mathbf e}} \def\bh{{\mathbf h}}
\def\bn{{\mathbf n}} \def\bu{{\mathbf u}}
\def\bx{{x}} \def\by{{y}} \def\bz{{z}}
\def\B1{{\mathbf 1}} \def\co{\complement}

\def\bmu{{\mbox{\boldmath${\mu}$}}}
\def\bnu{{\mbox{\boldmath${\nu}$}}}
\def\bPhi{{\mbox{\boldmath${\Phi}$}}}

\def\fA{{\mathfrak A}} \def\fB{{\mathfrak B}} \def\fC{{\mathfrak C}}
\def\fD{{\mathfrak D}} \def\fE{{\mathfrak E}} \def\fF{{\mathfrak F}}
\def\fW{{\mathfrak W}} \def\fX{{\mathfrak X}} \def\fY{{\mathfrak Y}}
\def\fZ{{\mathfrak Z}}

\def\rA{{\rm A}}  \def\rB{{\rm B}}  \def\rC{{\rm C}}
\def\rF{{\rm  F}} \def\rM{{\rm  M}}
\def\rS{{\rm S}}
\def\rT{{\rm T}}  \def\rW{{\rm W}}

\def\ov{\overline}  \def\wh{\widehat}  \def\wt{\widetilde}

\def\es {{\varnothing}}

\def\bt {\circ}
\def\cc {\circ}

\def\wt{\widetilde}

\def\wtm{{\wt m}} \def\wtn{{\wt n}} \def\wtk{{\wt k}}

\def\be{\begin{equation}}
\def\ee{\end{equation}}

\def\beal{\begin{array}{l}}
\def\beac{\begin{array}{c}}
\def\bear{\begin{array}{r}}
\def\beacl{\begin{array}{cl}}
\def\beall{\begin{array}{ll}}
\def\bealll{\begin{array}{lll}}
\def\beallll{\begin{array}{llll}}
\def\bealllll{\begin{array}{lllll}}
\def\beacr{\begin{array}{cr}}
\def\ena{\end{array}}

\def\bma{\begin{matrix}}
\def\ema{\end{matrix}}

\def\bpma{\begin{pmatrix}}
\def\epma{\end{pmatrix}}

\def\bcs{\begin{cases}}
\def\ecs{\end{cases}}

\def\diy{\displaystyle}

\def\sA{\mathscr A} \def\sB{\mathscr B} \def\sC{\mathscr C}
\def\sD{\mathscr D} \def\sE{\mathscr E}
\def\sF{\mathscr F} \def\sG{\mathscr G} \def\sI{\mathscr I}
\def\sL{\mathscr L} \def\sM{\mathscr M}
\def\sN{\mathscr N} \def\sO{\mathscr O}
\def\sP{\mathscr P} \def\sR{\mathscr R}
\def\sS{\mathscr S} \def\sS{\mathscr T}
\def\sU{\mathscr U} \def\sV{\mathscr V} \def\sW{\mathscr W}
\def\sX{\mathscr X} \def\sY{\mathscr Y} \def\sZ{\mathscr Z}

\def\BZ{{\mathbf Z}}

\def\D{D}

\def\bs {\overline \phi}
\def\bbLv {{\cal E}}

\def\iy{\infty}
\def\ct{\cdot}
\def\cl{\centerline}

\def\bbL{{\mathbb L}} \def\Pf{{\mathbf Z}} \def\f{{\vphi}} \def\g{{\Gam}} \def\s{{\phi}}
\def\boeta{{\mbox{\boldmath$\eta$}}}
\def\sq{\square} \def\tr{\triangle} \def\lz{\lozenge}
\def\cre{\color{red}} \def\cbl{\color{blue}}

\def\bmu{{\mbox{\boldmath${\mu}$}}} 

\def\bbA{{\mathbb A}} \def\bbE{{\mathbb E}} \def\bbH{{\mathbb H}} \def\bbN{{\mathbb N}}
\def\bbR{{\mathbb R}} \def\bbV{{\mathbb V}}  \def\bbW{{\mathbb W}} \def\bbZ{{\mathbb Z}}

\def\cA{{\mathcal A}} \def\cC{{\mathcal C}} \def\cE{{\mathcal E}} \def\cG{{\mathcal G}} 
\def\cN{{\mathcal N}} \def\cP{{\mathcal P}}

\def\Gam{\Gamma} \def\Lam{\Lambda} 
\def\vphi{\varphi} \def\vrho{\varrho}
\def\veps{\varepsilon}
\def\tD{{\tt D}} \def\tR{{\tt R}}
\def\tT{{\tt T}}
\def\wt{\widetilde} \def\ov{\overline}
\def\rd{{\rm d}} \def\rS{{\rm S}} 
\def\rExt{\rm{Ext}}
\def\rInt{\rm{Int}} \def\rSupp{\rm{Supp}}


\makeatletter
 \def\fps@figure{htbp}
\makeatother

\title{{\bf The hard-core model on planar lattices:\\ the disk-packing problem and\\ high-density phases}}

\author{\bf A. Mazel$^1$, I. Stuhl$^2$, Y. Suhov$^{2,3}$}

\date{}
\footnotetext{2010 {\em Mathematics Subject Classification:\; primary 60G60, 82B20, 82B26, 52C15}}

\footnotetext{{\em Key words and phrases:} hard-core model, exclusion distance, Gibbs
distributions, high-density/large fugacity, unit planar lattices, Pirogov--Sinai theory, 
dense-packing of disks, sliding

\quad $^1$ \footnotesize{AMC Health, New York, NY, USA;} $^2$ \footnotesize{Math Department, Penn State University, 
PA, USA;} $^3$ \footnotesize{DPMMS, University of Cambridge and St John's College, Cambridge, UK}.
}
\maketitle

\begin{abstract}
We study dense packings of disks and related Gibbs distributions representing high-density phases 
in the hard-core model on unit triangular, honeycomb and square lattices. The model is characterized 
by a Euclidean exclusion distance $D>0$ and a value of fugacity $u>0$. We use the Pirogov-Sinai theory 
to study the Gibbs distributions for a general $D$ when $u$ is large: $u>u_0(D)$. For infinite 
sequences of values $D$ we describe a complete high-density phase diagram: it exhibits a multitude of 
co-existing pure phases, and their number grows as $O(D^2)$. For the remaining values of $D$, except 
for those with sliding, the number of co-existing pure phases is still of the form $E(D)\geq O(D^2)$; 
however, the exact identification of the pure phases requires an additional analysis. Such an analysis 
is performed for a number of typical examples, which involves computer-assisted proofs. Consequently, for all values $D>0$ where sliding does not occur, we establish the existence of a phase transition.

The crucial steps in the study are (i) the identification of periodic ground states and (ii) the verification of 
the Peierls bound. This is done by using connections with algebraic number theory. In particular, 
a complete list of so-called sliding values of $D$ has been specified. As a by-product, we 
solve the disk-packing problem on the lattices under consideration. The number and structure of 
maximally-dense packings depend on the disk-diameter $D$, unlike the case of $\mathbb{R}^2$.

All assertions have been proved rigorously \cite{MSS1, MSS2}, some of the proofs are 
computer-assisted. 
\end{abstract}

\section{Introduction}

{\bf 1.1.} In this paper we report some rigorous results about the spherical hard-core 
(H-C)$^{*)}$\footnote{$^{*)}$For the reader's convenience, we provide the list of used abbreviations: 
H-C hard-core, GD Gibbs distribution, P-S Pirogov-Sinai, PGS periodic ground state, EGD extreme Gibbs 
distribution, AC admissible configuration, MDA maximally-dense admissible.}
model on planar lattices where the admissible configurations are packings of hard disks of diameter $D$ 
representing particles with the Euclidean exclusion distance $D$ and with centers at lattice 
sites.$^{**)}$ \footnote{$^{**)}$The value $D$ gives the minimal allowed distance at which the centers 
of disks can be placed.} 
Rigorous proofs of these results are contained in \cite{MSS1}, \cite{MSS2}.

When the fugacity/acivity $u>0$ is small, the particle system is in a low-density/dis\-or\-dered phase; 
mathematically it means that in the thermodynamic limit the system has a unique {\it Gibbs distribution} (GD). 
An important question is how the system evolves when the density/fugacity increases, e.g., whether it 
undergoes phase transitions. In a high-density regime (where $u$ is large) the H-C model is intrinsically 
related to the optimal (i.e., maximally-dense) disk-packing problem on the underlying lattice. Here the 
system is expected to become ordered, i.e., to be in a crystalline/solid phase (one or several). We 
show that this fact holds true in the thermodynamic limit for all but finitely many explicitly enumerated values 
of $D$, provided that $u$ is large enough:  $u\in (u_0,\infty)$ where $u_0$ depends on $D$ and the type 
of the lattice under consideration. The excluded values of $D$ exhibit a phenomenon of sliding defined and 
described in section 3.3. For all non-sliding values of $D$ we give a description of the large-fugacity phase 
diagram establishing that the pure phases always emerge from optimal/maximally dense disk-packing 
configurations (ground states), but not necessarily from all of them. Such non-uniqueness of pure phases proves the existence of a phase transition for all non-sliding values of $D$ \cite{Ge}, \cite{GHM}. The structure of ground states is 
analyzed in number-theoretical and geometric terms. 

We consider unit triangular, honeycomb and square lattices, $\bbA_2$, $\bbH_2$ and $\bbZ^2$. It turns 
out that only {\it attainable} values of $D$ (or rather $D^2$) are of interest, i.e., those that can be realized 
on the corresponding lattice. On $\bbA_2$ and $\bbH_2$ the attainable values $D^2$ are of the form 
$D^2=a^2+b^2+ab$, while on $\bbZ^2$ they obey $D^2=a^2+b^2$, where $a, b$ are non-negative 
integers. (For a non-attainable disk diameter $D'$ one applies the results for the smallest attainable $D$ 
such that $D>D'$.)

A complete description of the large fugacity phase diagram requires the determination of all pure 
phases; mathematically it means an identification of {\it extreme Gibbs distributions} (EGDs). A 
popular tool here is the Pirogov-Sinai (P-S) theory \cite{PS}, \cite{Z1}, based on (i) a specification 
of {\it periodic ground states} (PGSs) and (ii) verification of the {\it Peierls bound}\, for the statistical 
weight of a deviation of an admissible  particle configuration from a PGS. For reviews of the P-S 
theory, cf. \cite{FrVe}, \cite {Si}, \cite{Sl}, \cite{Z2}; for a general view in the context of statistical 
mechanics, see \cite{Br}.

Lattice H-C models (with a variety of hard-core shapes) attracted a considerable interest in statistical 
mechanics, beginning with \cite{Bu}, \cite{Do}, \cite{GF}, \cite{Ga}. Thereupon, an extensive mathematical and 
physical bibliography has been generated (including experimental studies of specific materials); cf. \cite{AGMS} - \cite{Ba2}, \cite{DKBP}, \cite{FAL}, \cite{HePr}, \cite{JMTR}, \cite{JL1}, \cite{JL2},  \cite{NR1}, \cite{NR2},  
\cite{TF} and references therein.

The existence of multiple EGDs in a high-density regime has been rigorously proved (i) in \cite{Do} 
on $\bbZ^d$ for $D=\sqrt{2}$ (for all $d>1$), (ii) in \cite{HePr} on $\mathbb{A}_2$ for a family of 
values $D$ and (iii) for a class of H-C lattice particle systems in \cite{JL2}. In relation to the spherical 
H-C model on $\bbA_2$, $\bbH_2$ and $\bbZ^2$, our work includes these results as partial cases of 
a general theory. 
 
An exact solution for a spherical H-C model was obtained for $D^2=3$ on the triangular lattice $\bbA_2$ 
\cite{Ba1}, \cite{Ba2}.

For some of the sliding values of $D$ on $\bbZ^2$, paper \cite{NR2} states the existence of a columnar
order; for the sliding value $D=2$ the presence of a columnar order has been proven in \cite{HaPe}.

Spherical H-C models on two-dimensional lattices gained a growing popularity in the recent physical 
literature \cite{AGMS}, \cite{DKBP}, \cite{FAL}, \cite{JMTR}, \cite{NR1}, \cite{NR2},  
\cite{TF} and references therein. Most of these papers specify the model as $k$-nn or $k$-NN exclusion, indicating that a particle 
excludes all nearest-neighboring sites, $2$-nd nearest-neighboring sites, ..., $k$-th nearest-neighboring 
sites  (all in the planar Euclidean metric). The value $D$ used to specify the model in the current paper 
is the Euclidean distance to the $(k+1)$-st nearest-neighbors.  A subject of interest is that, depending on 
the lattice type and the value of $D$, one expects different forms of phase transitions when the particle 
density/fugacity/chemical potential in the system increases. A progress in simulation and analytical 
techniques led to predictions for critical points and types of the related phase transitions for some initial 
values of $k$. For H-C models on $\bbZ^2$ we refer to \cite{FAL}, \cite{NR1}, \cite{NR2},
on $\bbA_2$ to \cite{AGMS}, \cite{JMTR}, and on $\bbH_2$ to \cite{DKBP}, \cite{TF}. 

Our work considers all values of $k\in\bbN$ and analyzes the model in the thermodynamic limit. We
give a detailed description of optimal/maximally-dense packings and -- for the cases without sliding -- 
of their associated large-fugacity phases.  Our findings regarding the large-fugacity pure phases coincide 
with those obtained in \cite{AMMGPS}, \cite{DKBP}, \cite{JMTR}, \cite{NR1} and \cite{TF} for initial 
values of $k$ considered therein. However, a rigorous analysis of criticality  remains an open (and 
challenging) mathematical problem.
\vskip .5cm  

{\bf 1.2.} A brief summary of our results is as follows. On each of the lattices $\bbZ^2$, $\bbA_2$ 
and $\bbH_2$, the set of attainable values of $D$ is partitioned into three distinct groups (A)--(C) 
characterized by different structures of extreme Gibbs distributions. (A) For some explicitly determined 
infinite sequences of numbers $D$ we provide a complete picture of the set of the large fugacity EGDs. It 
yields a multitude of co-existing pure phases where every periodic ground state generates an EGD. The 
number $E(D)$ of the EGDs grows as $O(D^2)$. (B) For the remaining infinite sets of values $D$, except 
for 39 particular numbers on $\bbZ^2$  and 4 on $\bbH_2$, there is still a multitude of coexisting pure 
phases, with $E(D)$ growing at least as $O(D^2)$, but not all PGSs generate EGDs. Here the problem is 
reduced to an identification of {\it dominant} PGSs (see section 5.1). The corresponding 
complete analysis has been done in \cite{MSS1} for a selection of typical examples. (C) For the $43$ 
excluded values of $D$ the model exhibits a phenomenon of {\it sliding} (see section 3.3).  

For a value $D$ from group (A) there is a finite number of PGSs, and any two of them are taken to 
each other by  an underlying lattice symmetry. Consequently, if one of these PGSs generates an EGD then 
so does each of them, and all generated EGDs are distinct. For a value $D$ from group (B) there are at 
least two symmetry classes of PGSs. PGSs from a given class are taken into each other by a lattice 
symmetry, but any two representatives of different symmetry classes are not lattice-symmetric to each 
other. Typically, only PGSs from one symmetry class generate EGDs;  for that reason such a class and 
the PGSs in it are called dominant.  For values $D$ from group C there is an infinite degeneracy of PGSs; 
typically these PGSs are layered configurations consisting of 1-dimensional layers. Moreover, each of these 
layers can be placed in at least two different positions independently of the positions of other layers, and all 
obtained configurations remain ground states. This represents a phenomenon of sliding in the 
two-dimensional H-C models under consideration. Note that a phenomenon of sliding exists in H-C models 
in higher dimensions as well although its precise definition is more involved.

\vskip .5cm

{\bf 1.3.} In the course of identifying ground states we solve the disk-packing problem on $\bbA_2, \bbH_2$ 
and $\bbZ^2$. Namely, given an attainable $D^2$, we establish the supremum of the disk-packing density 
among all packings/admissible configurations, both periodic and non-periodic, and show that it is 
achievable. 

Let $\mathbb{W}=\bbA_2, \bbH_2, \bbZ^2$ and $\Lambda_l$ be a square of side-length $l$ on $\bbR^2$. 
We define the maximal disk-packing density by
$$\delta(D, \bbW):=\sup_{\Phi}\Bigg\{\limsup_{l\to \infty} \frac{{\rm Area}(\Phi \cap \Lambda_l)}{
{\rm Area}(\Lambda_l)}\Bigg\},$$
where $\Phi$ is a packing of disks of diameter $D$ with centers at sites of $\bbW$.

It turns out that for any attainable {$D^2$} on $\bbA_2$ and any attainable $D^2$ on $\bbH_2$ divisible 
by $3$,
$$\delta(D, \bbA_2)=\delta(D, \bbH_2)=\frac{\pi}{2\sqrt{3}},$$
which gives the maximal disk-packing density on $\bbR^2$. Dense-packings on $\bbA_2$ have been 
considered in \cite{CD} as well. 

Next, for all attainable {$D^2$ on $\bbH_2$ non-divisible by $3$, except for} $D^2$ from the collection 
$\cN:=\{1, 4, 7, 13, 16, 28, 31, 49, 64, 67, 97, 133, 157, 256 \}$, the maximal disk-packing density has 
the form  
$$\delta(D, \bbH_2)=\frac{\pi D^2}{2\sqrt{3}(D^{\ast})^2}.$$
Here $D^*>D$ is the closest attainable value with $(D^*)^2$ divisible by $3$. Finally, for $D^2\in{\mathcal N}$, 
the identification of the dense-packings on $\bbH_2$  is done case-by-case, and we refer the reader 
to \cite{MSS1} for their detailed description. These 
configurations consist of periodically alternating triangular tiles of two distinct types, not necessarily forming a sub-lattice.
 
On the other hand, for all attainable $D$ on $\mathbb{Z}^2$ 
$$
\delta(D, \bbZ^2)=\frac{\pi D^2}{4S(D)},
$$
where $S(D)/2$ is the solution to optimization problem (5); see below.

Furthermore, for any attainable non-sliding $D$ on $\bbA_2, \bbH_2, \bbZ^2$ we describe all periodic 
optimizers, i.e., periodic packings achieving the maximal density.
On $\bbA_2$ these are $D$-sub-lattices and their shifts (see section 3.2), while on 
$\bbZ^2$ they 
are the so-called maximally-dense admissible sub-lattices and their shifts (see section 3.5).
On $\bbH_2$ the optimizers are $D$-sub-lattices and their shifts when 
$D^2$ is divisible by 3, and $D^*$-sub-lattices and their shifts when $D^2\not\in{\mathcal N}$
and $D^2$ is not divisible by 3 (see section 3.4). For {$D^2=1, 13, 16, 28, 49, 64, 
67, 97, 157, 256$ on $\bbH_2$ there are finitely many optimal periodic packings but they 
are not necessarily sub-lattices: cf. \cite{MSS1}. }
For the $43$ sliding values $D$ (on $\bbZ^2$ and $\bbH_2$), we construct infinitely many 
optimal periodic packings. 

We would like to note that, for each of $\bbA_2$,  $\bbZ^2$ and $\bbH_2$, 
there are infinitely many values of $D$ for which the optimizer is unique up to lattice 
symmetries, and there are also infinitely many values of $D$ with multiple,
but finitely many, optimizers up to lattice symmetries. Moreover, these alternatives exhaust
all attainable non-sliding values of $D$.
The number of optimal packings depends upon $D$; on $\bbZ^2$ and $\bbH_2$ the disks in 
these packings do not necessarily touch their `neighbors'.

\section{Gibbs distributions} 

{\bf 2.1.} A
$D$-{\it admissible configuration} ($D$-AC) on $\bbW$ is a map $\phi:\bbW\to \{0, 1\}$ such that
${\rm{dist}}(x,y)\geq D$, for all $x, y\in \bbW$ with $\phi(x)=\phi(y)=1$. Sites $x\in \bbW$ with 
$\phi(x)=1$ are treated as occupied, those with $\phi(x)=0$ as vacant. The set of $D$-ACs is 
denoted by $\cA=\cA(D,\bbW)$. Similar definitions can be introduced for a subset
$\bbV\subseteq\bbW$. Cf. Fig. 1. Given $\bbV\subseteq\bbW$, we say that configuration 
$\psi\in\cA(D,\bbV$) is compatible with $\phi\in\cA$ if the concatenation 
$\psi \vee (\phi\hskip -3pt\upharpoonright_{\bbW\setminus\bbV})\in\cA$. Define the probability distribution
$\mu_{\bbV} (\, \cdot \, || \phi)$ on $\cA(D,\bbV$): 
$$\mu_{\bbV} (\psi || \phi):= \begin{cases}\dfrac{u^{\sharp \psi}}{Z(\bbV||\phi)}, &
\hbox{$\psi$\, is\, compatible\, with\, $\phi$,}\\
0, & \hbox{otherwise.}\end{cases}\eqno (1)$$
Here $\sharp \psi$ stands for the number of occupied sites in $\psi\in\cA(D,\bbV)$ and 
$Z(\mathbb{V} || \phi)$ is the partition function in $\mathbb{V}$ with the boundary condition 
$\phi\in \mathcal{A}$: 
$$Z(\mathbb{V} || \phi):=\sum\limits_{\psi\in
\mathcal{A}(D,\mathbb{V})} u^{\sharp
\psi}\mathbf{1}(\psi\,\,
\hbox{is compatible with}\,\, \phi).\eqno (2)$$ 
$\mu_{\mathbb{V}} (\, \cdot \, || \phi)$ is called an H-C Gibbs distribution in `volume' ${\mathbb V}$ with the
boundary condition $\phi$ at fugacity $u>0$. 

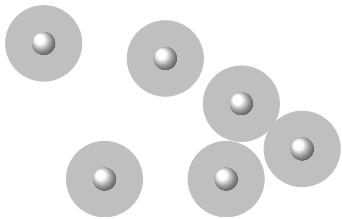
\begin{figure} \label{D-AC D^2=8} 
\begin{center}
\begin{tikzpicture}[scale=0.2]
\clip (-6.0, -6.5) rectangle (17.2, 7.6);

\filldraw[lightgray] (5,4) circle (2.5); 
\filldraw[lightgray] (-3,5) circle (2.5);  
\filldraw[lightgray] (10,1) circle (2.5); 
\filldraw[lightgray] (9,-4) circle (2.5); 
\filldraw[lightgray] (14,-2) circle (2.5);
\filldraw[lightgray] (1,-4) circle (2.5);

\foreach \pos in {(5,4),(-3,5),(10,1),(9,-4), (14,-2),(1,-4)}
\shade[shading=ball, ball color=white] \pos circle (0.75);

\end{tikzpicture} \end{center}

\caption{An admissible configuration}
\end{figure}

{\bf 2.2.} A modern mathematical definition describes a  Gibbs distribution $\bmu$ on $\bbW$ as a
(Borel) probability measure concentrated on set $\cA$ and related to a (Gibbsian) {\it specification} 
formed by the family $\{\mu_{\bbV_n}(\,\cdot\,||\phi)\}$. Cf. \cite{Ge}, Chapter 2, particularly, 
Definition 2.9. Of a particular importance are extreme Gibbs distributions which cannot be written as 
a non-trivial mixtures of other GDs. In this paper we work with GDs obtained as thermodynamic limits
in  the Van Hove sense
$$\lim_{\bbV_n\nearrow\bbW}\mu_{\bbV_n}(\,\cdot\,||\phi)\eqno (3)$$
in the weak topology. If it exists, the limiting measure in (3) is denoted by $\bmu_\phi$, and we say that 
$\bmu_\phi$ is generated by $\phi\in\cA$.


This work studies the H-C EGDs for values $u$ large enough; such an assumption is in force throughout
the paper and not stressed every time again. The structure of set $\cE (D)$ depends both on the choice 
of lattice $\bbW$ and arithmetic properties of $D$ and is studied using the P-S theory based on the notion 
of a PGS. 

A ground state is defined as a $D$-AC $\vphi\in \cA$  with the property that one cannot remove finitely 
many particles from $\vphi$ and replace them by a larger number of particles without breaking admissibility. 
A PGS is a ground state invariant under two non-collinear lattice shifts, hence under their convolutions. 
A parallelogram spanned by the corresponding vectors is called a period of a PGS. We let 
$\cP (D)=\cP (D,\bbW)$ denote the set of PGSs for given $D$ and $\bbW$.

The P-S theory \cite{PS}, \cite{Z1}, states that, under certain assumptions, (a) every 
periodic EGD $\bmu$ is generated by a PGS $\vphi$ in the sense of (3): $\bmu=\bmu_\vphi$, 
(b) every periodic EGD $\bmu$ has a list of properties stressing its pure-phase character 
(see items {\bf{(P1)}}--{\bf{(P5)}} in section 5.1.), (c) there are no other 
periodic EGDs. Moreover, general results from \cite{DS} ensure that in a wide class of 
two-dimensional models, including the H-C model, there is no non-periodic EGD at large 
fugacities, i.e., non-periodic ground states do not generate EGDs. Hence, indentifying the PGSs
and applying the above-mentioned general results from \cite{PS}, \cite{Z1}, \cite{DS}, including 
the identification of dominant/stable PGSs, would allow one to describe all large-fugacity EGDs. 
The assumptions from  \cite{PS}, \cite{Z1}, \cite{DS} that we need to verify are that {\bf{(I)}} 
there are finitely many PGSs, and {\bf{(II)}} a deviation from a PGS is controlled by a suitable 
Peierls bound based on an appropriate notion of a {\it contour}. Verifying these assumptions
is a key part of this work.


\def\rr{0.86602540378443864676372317075294} 

\def\hexagongrid{
\draw [yscale=sqrt(3/4), xslant=0.5] (-\n, -\n) grid (\n, \n);
\draw [yscale=sqrt(3/4), xslant=-0.5] (-\n, -\n) grid (\n, \n);
}

\def\rectangulargrid{
\clip[yscale=sqrt(3/4)] (-\n, -\n) rectangle (\n, \n);
\draw [yscale=sqrt(3/4), xslant=0.5] (-2 *\n, -\n) grid (2 *\n, \n);
\draw [yscale=sqrt(3/4), xslant=-0.5] (-2 *\n, -\n) grid (2 *\n, \n);
}

\def\sublattice{
\clip[yscale=sqrt(3/4), xslant=0.5] (0, \n+\nn) -- (\n+\nn, 0) -- (\n+\nn, -\n-\nn) -- (0, -\n-\nn) -- (-\n-\nn, 0) -- (-\n-\nn, \n+\nn) -- cycle;
\foreach \x in {-\n,...,\n}
 \foreach \y in {-\n,...,\n}
 {
  \def\xx{\x *\aa + 0.5 *\x *\bb - 0.5 *\y *\bb + 0.5 *\y *\aa}
  \def\yy{\rr *\x *\bb + \rr *\y *\aa + \rr *\y *\bb}

  \shade[shading=ball, ball color=white] (\xx, \yy) circle (.2);
 }
}

\def\sublatticeinrectangle{
\clip[yscale=sqrt(3/4)] (-\nn, -\nn) rectangle (\nn, \nn);
\foreach \x in {-\n,...,\n}
 \foreach \y in {-\n,...,\n}
 {
  \def\xx{\x *\aa + 0.5 *\x *\bb - 0.5 *\y *\bb + 0.5 *\y *\aa}
  \def\yy{\rr *\x *\bb + \rr *\y *\aa + \rr *\y *\bb}

  \shade[shading=ball, ball color=white] (\xx + \dx + 0.5 *\dy, \yy + \rr *\dy) circle (.3);
 }
}

\def\inclinedgrid{
	\draw [yscale=\bb * sqrt(3/4), xslant=\aa + 0.5 *\bb, xscale=\aa *\aa / \bb + \aa  + \bb , yslant=-\aa / \bb, ultra thick] (-\n, -\n) grid (\n, \n);
}

\def\rectsublattice{
\clip[yscale=sqrt(3/4)] (-\nn, -\nn) rectangle (\nn, \nn);
\foreach \x in {-\n,...,\n}
 \foreach \y in {-\n,...,\n}
 {\filldraw[yscale=sqrt(3/4), xslant=0.5, \cc] (\ss * \x + \dx, \ss * \y + \dy) rectangle (\ss * \x + \ss - 1 + \dx, \ss * \y + \ss - 1 + \dy);}
\definecolor{gray1}{gray}{0.1}
\definecolor{gray2}{gray}{0.2}
\definecolor{gray3}{gray}{0.3}
\definecolor{gray4}{gray}{0.4}
\definecolor{gray5}{gray}{0.5}
\definecolor{gray6}{gray}{0.6}
\definecolor{gray7}{gray}{0.7}
\definecolor{gray8}{gray}{0.8}
\definecolor{gray9}{gray}{0.9}
\clip[yscale=sqrt(3/4)] (-\nn, -\nn) rectangle (\nn, \nn);
\foreach \x in {-\n,...,\n}
 \foreach \y in {-\n,...,\n}
 {\filldraw[yscale=sqrt(3/4), xslant=0.5, \cc] (\ss *\x + \dx, \ss *\y + \dy) rectangle (\ss *\x + \ss - 1 + \dx, \ss *\y + \ss - 1 + \dy);}}

\definecolor{gray1}{gray}{0.1}
\definecolor{gray2}{gray}{0.2}
\definecolor{gray3}{gray}{0.3}
\definecolor{gray4}{gray}{0.4}
\definecolor{gray5}{gray}{0.5}
\definecolor{gray6}{gray}{0.6}
\definecolor{gray7}{gray}{0.7}
\definecolor{gray8}{gray}{0.8}
\definecolor{gray9}{gray}{0.9}


\section{Periodic ground states} 

{\bf 3.1.} Let us consider assumption {\bf{(I)}}. It turns out 
that, apart from a few exceptional values of $D$ on $\bbH_2$ and $\bbZ^2$
which have to be analyzed separately, 
the PGSs on $\bbW=\bbA_2, \bbH_2, \bbZ^2$ are given by  
{\it maximally-dense admissible} (MDA) sub-lattices
$\bbE\subset\bbW$ and their $\bbW$-shifts. This is a non-trivial result established in \cite{MSS1, 
MSS2}. In the simplest cases an MDA sub-lattice arises because an admissible triangle of minimal 
area uniquely tessellates the corresponding lattice, up to lattice symmetries and shifts. (Cf. Theorem 
I  in \cite{MSS1}, Lemma I and Theorem 1 (i) in \cite{MSS2}). A non-unique tessellation by a given minimal area triangle always 
leads to sliding (section 2.2 in \cite{MSS2}, section 8 in \cite{MSS1}). 

Furthermore, on $\bbH_2$ there are cases where no admissible triangle of minimal area tessellates 
the whole lattice. In the majority of these cases (constituting an infinite amount of values of $D$), 
an additional number-theoretical argument verifies that PGSs are still given by MDA sub-lattices 
(Theorem 1 (ii), Lemma 4.10 (cases $D^2=64, 67$) in \cite{MSS1}). Among the remaining cases 
there are values of $D$ for which some of the corresponding PGSs are not  MDA sub-lattices. (Theorem 
1 (ii), Lemma 4.10 (all cases except for $D^2=64$)).

Observe that in all cases one can construct non-periodic ground states, e.g., by filling two half-planes with different PGSs. Additional clarifications are provided in section 3.6.

The $\bbW$-{\it shifts} and $\bbW$-{\it symmetries} ($\bbW$-reflections/rotations) define a partition
of set \\ $\cP (D,\bbW )$ into PGS-{\it equivalence classes}. Let $K=K(D,\bbW )$ denote the number of
PGS-equivalence classes. Recall that, by the definition of sliding, if $D$ is non-sliding then $K<\infty$. For 
a case when $D$ is non-sliding and on $\bbW=\bbH_2$ when $D^2\not\in{\mathcal N}$ a PGS is always 
an MDA sub-lattice or its shift.  Let $\sigma$ stand for the number of lattices sites in the half-open 
fundamental parallelogram of an MDA-sub-lattice and $m_j$ denote the number of distinct 
MDA-sub-lattices within a given PGS-equivalence class, for $j=1,\ldots ,K$. Then, for the cardinality 
$\sharp\cP (D,\bbW)$ of the set of PGSs $\cP (D,\bbW)$ we have  
$$\sharp\cP (D, \bbW)=\sigma\sum\limits_{j=1}^{K}m_j.\eqno (4)$$

Further specifications for $K$, $\sigma$ and $m_j$ depend on the choice of $\bbW$ and $D$ and are 
provided below. In particular, dependence upon $D$ is non-monotonic and exhibits connections with 
the algebraic number theory.
\vskip .5cm

{\bf 3.2.} On $\bbA_2$, for any $D$ every MDA sub-lattice is obviously a $D$-sub-lattice, 
for which a fundamental parallelogram is formed by two equilateral triangles with side-length $D$ 
($D$-triangles). This is due to the fact that an equilateral triangular arrangement solves the 
dense-packing problem in $\bbR^2$. Consequently, on $\bbA_2$ the value $\sigma =D^2$.

It is not hard to see that on $\bbA_2$, admissibility of $D$ is equivalent 
to solvability of the Diophantine equation $D^2=a^2+b^2+ab$, $a,b\in\bbZ$  (i.e., to the fact that $D^2$ 
is a L\"oschian number). Therefore, the number $K$ of PGS-classes in Eqn (4) is directly related to the 
number of solutions $a,b\in\bbZ$, $0\leq a\leq b$, to this equation. It is well-known that a solution exists 
iff the prime decomposition of $D^2$ contains primes of the from $3v+2$ in even powers. See Proposition 9.1.4, \cite{IR}.

We will consider three disjoint groups of values of $D$ (or
$D^2$) deemed TA1, TA2 and TB, and based on further conditions
upon prime factors of $D^2$. The first character in the labels 
TA1, TA2 and TB reflects the choice of a lattice (T stands for
triangular). Derivation of these results based on known number-theoretical facts can be found in Theorems 1 - 3 from \cite{MSS1}.

First, take the case TA1: here $D^2$ is of the form $D^2=a^2$ or
$D^2=3a^2$ where $a\in \bbN$ has only prime factors 3 or prime factors of
the form $3v+2$, i.e., the prime decomposition of $D^2$ 
has no primes of form $3v+1$. Then the above Diophantine equation has a
unique solution, and the $D$-sub-lattice is $\bbA_2$-reflection invariant;
consequently, $K=1$, $m_1=1$ and the number of PGSs is $D^2$.  
{\begin{figure}[H]\begin{center}
(a) \begin{tikzpicture}[scale=0.48]
\clip (2, -0.5) rectangle (14.4, 8.1);
\draw[yscale=sqrt(3/4), xslant=0.5] (-4,-2) grid (29, 12);
\draw[yscale=sqrt(3/4), xslant=-0.5] (-1,-2) grid (29, 12);

\draw [line width=0.55mm] (4.5,3*\rr)--(7.5,3*\rr)--(6,6*\rr)--(4.5,3*\rr);
\draw [line width=0.55mm] (6,6*\rr)--(9,6*\rr)--(7.5,3*\rr);
\draw [line width=0.55mm] (9,6*\rr)--(10.5,3*\rr)--(9,0)--(7.5,3*\rr)
--(10.5, 3*\rr);
\foreach \pos in {(3,0), (6,0), (9, 0), (12, 0), (15, 0), 
(4.5, 3*\rr), (7.5, 3*\rr), (10.5, 3*\rr), (13.5, 3*\rr), (16.5, 3*\rr), 
(3,6*\rr),(6,6*\rr),(9,6*\rr),(12,6*\rr),(15,6*\rr),
(4.5,9*\rr), (7.5, 9*\rr), (10.5, 9*\rr), (13.5, 9*\rr), (16.5, 9*\rr),
(3,12*\rr),(6,12*\rr),(9,12*\rr),(12,12*\rr),(15,12*\rr)}
\shade[shading=ball, ball color=lightgray] \pos circle (.33);
\end{tikzpicture}\qquad (b) \begin{tikzpicture}[scale=0.48]
\clip (2,-0.5) rectangle (14.4,8.1);
\draw[yscale=sqrt(3/4), xslant=0.5] (-4,-2) grid (29, 12);
\draw[yscale=sqrt(3/4), xslant=-0.5] (-1,-2) grid (29, 12);
\draw [line width=0.55mm] (4,4*\rr)--(7.5,5*\rr)--(6.5,1*\rr)--(4,4*\rr);
\draw [line width=0.55mm] (4,4*\rr)--(5,8*\rr)--(8.5,9*\rr)--(7.5,5*\rr)
--(5,8*\rr);
\draw [line width=0.55mm] (8.5,9*\rr)--(11,6*\rr)--(7.5,5*\rr);
\foreach \pos in {(3, 0), (6.5,1*\rr), (10,2*\rr), (13.5,3*\rr), (17,4*\rr),
(4,4*\rr),(7.5,5*\rr),(11,6*\rr),(14.5,7*\rr),
(5,8*\rr),(8.5,9*\rr),(12,10*\rr),(15.5,11*\rr),
(2.5,11*\rr),(6,12*\rr)}
\shade[shading=ball, ball color=lightgray] \pos circle (.33);
\end{tikzpicture}
\caption{PGSs on ${\mathbb A}_2$,  
for $D^2=$ 9 (Case TA1) (a), and  $D^2=$ 13 (Case TA2) (b).
The number of PGSs is 9  and 26, respectively.}
\end{center}
\end{figure}
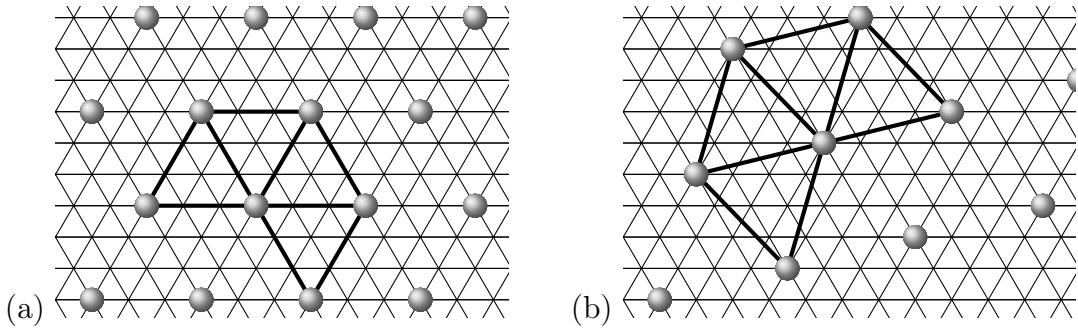} 

Case TA2 emerges when the equation has a unique solution such
that $a< b$, with $a, b\in \bbN$. A value $D^2$ belongs to the TA2 case iff
its prime decomposition has a single prime of form $3v+1$. In this case
$K=1$, the $D$-sub-lattice is not 
$\bbA_2$-reflection invariant, $m_1=2$, and the number of
PGSs is $2D^2$. Cf. Fig. 2.

 The 3rd case, TB, covers all remaining values of $D$; in this
case the equation has multiple solutions. Each solution $a,b$,
$0\leq a\leq b$, generates a PGS-equivalence class, and its
cardinality is $D^2$ or $2D^2$, similarly to the above cases 
TA1 and TA2. Cf. Fig. 3.

{\begin{figure} \begin{center}
(a) \begin{tikzpicture}[scale=0.32]
\clip (2, -0.5) rectangle (18.4, 15.2);

\draw[yscale=sqrt(3/4), xslant=0.5] (-9,-2) grid (39, 21);
\draw[yscale=sqrt(3/4), xslant=-0.5] (-9,-2) grid (49, 21);

\draw [line width=0.6mm] (10,0)--(6.5, 7*\rr)--(13.5, 7*\rr)
--(10,0)--(17,0)--(13.5, 7*\rr);
\draw [line width=0.6mm] (6.5, 7*\rr)--(10,14*\rr)--(17,14*\rr)
--(13.5, 7*\rr)--(10,14*\rr);
\foreach \pos in {(3,0),(10,0),(17,0),(24,0), 
(6.5, 7*\rr), (13.5, 7*\rr), (20.5, 7*\rr), 
(3,14*\rr),(10,14*\rr),(17,14*\rr),(24,14*\rr),
(6.5,21*\rr), (13.5,21*\rr), (20.5,21*\rr)}
\shade[shading=ball, ball color=lightgray] \pos circle (.45);
\end{tikzpicture}\,\,\qquad (b)\begin{tikzpicture}[scale=0.32]
\clip (2, 1.5) rectangle (18.4, 17.2);

\draw[yscale=sqrt(3/4), xslant=0.5] (-9,-2) grid (39, 21);
\draw[yscale=sqrt(3/4), xslant=-0.5] (-9,-2) grid (49, 21);

\draw [line width=0.6mm] (4,8*\rr)--(5,16*\rr)--(11.5,19*\rr)
--(10.5,11*\rr)--(5,16*\rr);
\draw [line width=0.6mm] (9.5,3*\rr)--(4,8*\rr)--(10.5,11*\rr)
--(16,6*\rr)--(9.5,3*\rr)--(10.5,11*\rr);
\foreach \pos in {(3,0),(9.5,3*\rr),(16,6*\rr),(22.5,9*\rr),
(4,8*\rr),(10.5,11*\rr),(17,14*\rr),(23.5,17*\rr),
(5,16*\rr),(11.5,19*\rr)}
\shade[shading=ball, ball color=lightgray] \pos circle (.45);
\end{tikzpicture}
\caption{PGSs on ${\mathbb A}_2$ for $D^2=49$ (Case  TB).
There are 49 horizontal PGSs (a) and 98 inclined (b). The horizontal 
PGSs are the only dominant ones.}
\end{center}
\end{figure}
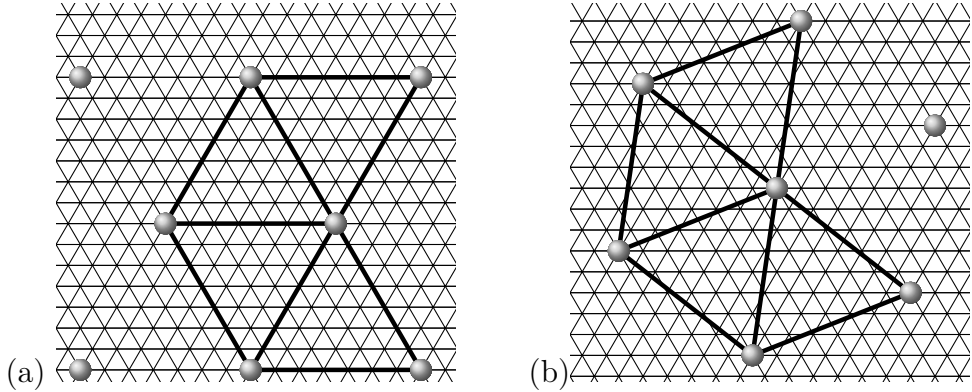} 
Classes TA1, TA2, TB form a partition of values of $D$ on $\mathbb{A}_2$. Given $D$ the knowledge of the corresponding class provides a specification of the number of PGSs and their structure.


\vskip .5cm

{\bf 3.3.} Now consider $\bbW=\bbH_2, \bbZ^2$. As was said, for some values $D^2$ on $\bbH_2$ and
$\bbZ^2$ we encounter a phenomenon of sliding. It occurs when one
can pass from one PGS to another without any local
loss in `energy' i.e., without decreasing the local particle numbers.

{\begin{figure}
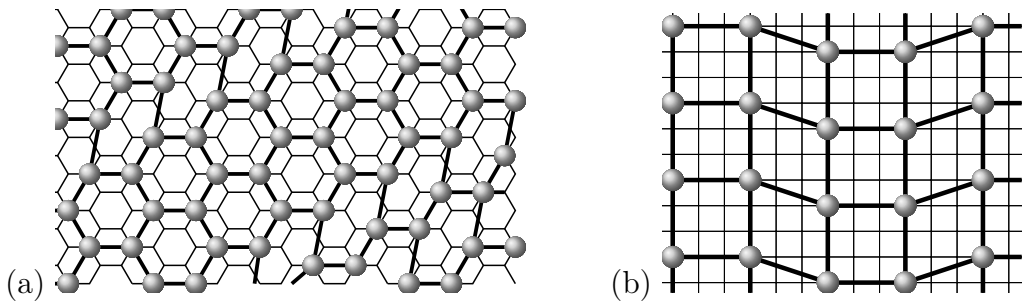

\begin{center}
(a) 


\caption{Sliding on ${\mathbb H}_2$ for $D^2=4$ (a) and on ${\mathbb Z}^2$
for $D^2=9$ (b).} 
\end{center}
\end{figure}} 

 
On $\bbH_2$ there are just 4 sliding values: $D^2=$ 4, 7, 31, 133; the
proof of this fact is computer-assisted \cite{MSS1}. Cf. Fig. 4 (a). 

On  $\bbZ^2$ there exist 39 sliding values: $D^2=$ 4, 8, 9, 18,
20, 29, 45, 72, 80, 90, 106, 121, 157, 160, 218, 281, 392,  521,
698, 821, 1042, 1325, 1348, 1517, 1565, 2005, 2792, 3034, 3709, 
4453, 4756, 6865, 11449, 12740, 13225, 15488, 22784, 29890, 
37970. Cf. Fig. 4 (b). This was first conjectured by the authors \cite{MSS2}, then in
\cite{K1} it was proved that the sliding list is finite, then finally
the completeness of the above list was independently established
in \cite{MSS2}, \cite{K2}. A part of the list was identified in \cite{NR2} as candidates for occurrence of a columnar order. In a recent paper \cite{HaPe}, the case of $D^2=4$ on $\bbZ^2$ was 
rigorously analyzed, confirming existence of four EGDs with a columnar order.

In what follows we consider the non-sliding values of $D$ only.\vskip .5cm

{\bf 3.4.} Additionally, on $\bbH_2$ there are exceptional values $D^2=$1, 13,
16, 28, 49, 64, 97, 157, 256 where the PGS-class is unique but is 
non-lattice. Cf. Fig. 5. Furthermore, for $D^2=$ 67 there are two classes 
one of which is non-lattice. These values are analyzed via a
special approach (see \cite{MSS1}, Theorems 12, 13) and not discussed in this article.

{\begin{figure}
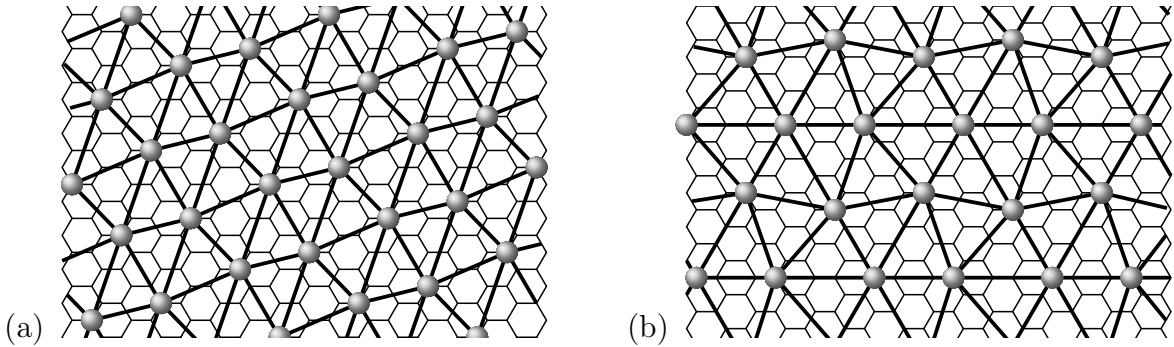
 
\begin{center}
(a)
 \end{center}
\caption{Non-lattice PGSs on ${\mathbb H}_2$ for $D^2=$ 13 (a)
and $D^2=$ 16 (b).}
\end{figure}} 


Thus, in what follows, on $\bbH_2$ we refer to non-exceptional values of $D$: 
here every PGS is an MDA-sub-lattice. More precisely,
if $3$ divides $D^2$ then every PGS is a $D$-sub-lattice since a
$D$-triangle can be inscribed in $\bbH_2$. Accordingly, we 
define groups of values $D^2$ deemed HA1, HA2 and HB and formed 
by the values from cases TA1, TA2 and TB divisible by 3.
When $D^2$ belongs to one of these H-cases, the theory goes in
parallel to the respective T-case. See Theorems I (ii), 7, 8, 9 in \cite{MSS1}. In particular, this yields 
the same values $K$, $m_k$ while $\sigma=2D^2/3$. Cf. Fig. 6.

\begin{figure}\begin{center} 
(a) \begin{tikzpicture}[scale=0.29]
\clip (-0.9, -1.2*\rr) rectangle (15.5, 14*\rr);

\draw [line width=0.2mm] (0.5,1*\rr) -- (1.5,1*\rr) -- (2,0) -- (1.5,-1*\rr)
-- (0.5,-1*\rr) -- (0,0) -- (0.5,1*\rr);
\draw [line width=0.2mm] (0.5,1*\rr) -- (0,2*\rr) -- (0.5,3*\rr) -- (1.5,3*\rr) 
-- (2,2*\rr) -- (1.5,1*\rr);
\draw [line width=0.2mm] (0.5,3*\rr) -- (0,4*\rr) -- (0.5,5*\rr) -- (1.5,5*\rr) 
-- (2,4*\rr) -- (1.5,3*\rr);
\draw [line width=0.2mm] (0.5,5*\rr) -- (0,6*\rr) -- (0.5,7*\rr) -- (1.5,7*\rr) 
-- (2,6*\rr) -- (1.5,5*\rr);
\draw [line width=0.2mm] (0.5,7*\rr) -- (0,8*\rr) -- (0.5,9*\rr) -- (1.5,9*\rr) 
-- (2,8*\rr) -- (1.5,7*\rr);
\draw [line width=0.2mm] (0.5,9*\rr) -- (0,10*\rr) -- (0.5,11*\rr) -- (1.5,11*\rr) 
-- (2,10*\rr) -- (1.5,9*\rr);
\draw [line width=0.2mm] (0.5,11*\rr) -- (0,12*\rr) -- (0.5,13*\rr) -- (1.5,13*\rr) 
-- (2,12*\rr) -- (1.5,11*\rr);
\draw [line width=0.2mm] (0.5,13*\rr) -- (0,14*\rr) -- (0.5,15*\rr) -- (1.5,15*\rr) 
-- (2,14*\rr) -- (1.5,13*\rr);
\draw [line width=0.2mm] (0.5,15*\rr) -- (0,16*\rr) -- (0.5,17*\rr) -- (1.5,17*\rr) 
-- (2,16*\rr) -- (1.5,15*\rr);
\draw [line width=0.2mm] (0.5,17*\rr) -- (0,18*\rr) -- (0.5,19*\rr) -- (1.5,19*\rr) 
-- (2,18*\rr) -- (1.5,17*\rr);
\draw [line width=0.2mm] (0.5,19*\rr) -- (0,20*\rr) -- (0.5,21*\rr) -- (1.5,21*\rr) 
-- (2,20*\rr) -- (1.5,19*\rr);

\draw [line width=0.2mm]  (3,0) -- (3.5,-1*\rr);
\draw [line width=0.2mm] (2,2*\rr) -- (3,2*\rr) -- (3.5,1*\rr) -- (3,0) -- (2,0);
\draw [line width=0.2mm] (2,4*\rr) -- (3,4*\rr) -- (3.5,3*\rr) -- (3,2*\rr);
\draw [line width=0.2mm] (2,6*\rr) -- (3,6*\rr) -- (3.5,5*\rr) -- (3,4*\rr);
\draw [line width=0.2mm] (2,8*\rr) -- (3,8*\rr) -- (3.5,7*\rr) -- (3,6*\rr);
\draw [line width=0.2mm] (2,10*\rr) -- (3,10*\rr) -- (3.5,9*\rr) -- (3,8*\rr);
\draw [line width=0.2mm] (2,12*\rr) -- (3,12*\rr) -- (3.5,11*\rr) -- (3,10*\rr);
\draw [line width=0.2mm] (2,14*\rr) -- (3,14*\rr) -- (3.5,13*\rr) -- (3,12*\rr);
\draw [line width=0.2mm] (2,16*\rr) -- (3,16*\rr) -- (3.5,15*\rr) -- (3,14*\rr);
\draw [line width=0.2mm] (2,18*\rr) -- (3,18*\rr) -- (3.5,17*\rr) -- (3,16*\rr);
\draw [line width=0.2mm] (2,20*\rr) -- (3,20*\rr) -- (3.5,19*\rr) -- (3,18*\rr);
\draw [line width=0.2mm] (3.5,21*\rr) -- (3,20*\rr);

\draw [line width=0.2mm]  (3.5,1*\rr) -- (4.5,1*\rr) -- (5,0) -- (4.5,-1*\rr)
--(3.5,-1*\rr);
\draw [line width=0.2mm]  (3.5,3*\rr) -- (4.5,3*\rr) -- (5,2*\rr) -- (4.5,1*\rr);
\draw [line width=0.2mm]  (3.5,5*\rr) -- (4.5,5*\rr) -- (5,4*\rr) -- (4.5,3*\rr);
\draw [line width=0.2mm]  (3.5,7*\rr) -- (4.5,7*\rr) -- (5,6*\rr) -- (4.5,5*\rr);
\draw [line width=0.2mm]  (3.5,9*\rr) -- (4.5,9*\rr) -- (5,8*\rr) -- (4.5,7*\rr);
\draw [line width=0.2mm]  (3.5,11*\rr) -- (4.5,11*\rr) -- (5,10*\rr) -- (4.5,9*\rr);
\draw [line width=0.2mm]  (3.5,13*\rr) -- (4.5,13*\rr) -- (5,12*\rr) -- (4.5,11*\rr);
\draw [line width=0.2mm]  (3.5,15*\rr) -- (4.5,15*\rr) -- (5,14*\rr) -- (4.5,13*\rr);
\draw [line width=0.2mm]  (3.5,17*\rr) -- (4.5,17*\rr) -- (5,16*\rr) -- (4.5,15*\rr);
\draw [line width=0.2mm]  (3.5,19*\rr) -- (4.5,19*\rr) -- (5,18*\rr) -- (4.5,17*\rr);
\draw [line width=0.2mm]  (3.5,21*\rr) -- (4.5,21*\rr) -- (5,20*\rr) -- (4.5,19*\rr);

\draw [line width=0.2mm]  (5,0) -- (6,0)--(6.5,-1*\rr );
\draw [line width=0.2mm]  (5,2*\rr) -- (6,2*\rr)--(6.5,1*\rr )--(6,0);
\draw [line width=0.2mm]  (5,4*\rr) -- (6,4*\rr)--(6.5,3*\rr )--(6,2*\rr);
\draw [line width=0.2mm]  (5,6*\rr) -- (6,6*\rr)--(6.5,5*\rr )--(6,4*\rr);
\draw [line width=0.2mm]  (5,8*\rr) -- (6,8*\rr)--(6.5,7*\rr )--(6,6*\rr);
\draw [line width=0.2mm]  (5,10*\rr) -- (6,10*\rr)--(6.5,9*\rr )--(6,8*\rr);
\draw [line width=0.2mm]  (5,12*\rr) -- (6,12*\rr)--(6.5,11*\rr )--(6,10*\rr);
\draw [line width=0.2mm]  (5,14*\rr) -- (6,14*\rr)--(6.5,13*\rr )--(6,12*\rr);
\draw [line width=0.2mm]  (5,16*\rr) -- (6,16*\rr)--(6.5,15*\rr )--(6,14*\rr);
\draw [line width=0.2mm]  (5,18*\rr) -- (6,18*\rr)--(6.5,17*\rr )--(6,16*\rr);
\draw [line width=0.2mm]  (5,20*\rr) -- (6,20*\rr)--(6.5,19*\rr )--(6,18*\rr);
\draw [line width=0.2mm]  (6.5,21*\rr )--(6,20*\rr);

\draw [line width=0.2mm]  (6.5,1*\rr) -- (7.5,1*\rr)--(8,0)--(7.5,-1*\rr)
--(6.5,-1*\rr);
\draw [line width=0.2mm]  (6.5,3*\rr) -- (7.5,3*\rr)--(8,2*\rr)--(7.5,1*\rr);
\draw [line width=0.2mm]  (6.5,5*\rr) -- (7.5,5*\rr)--(8,4*\rr)--(7.5,3*\rr);
\draw [line width=0.2mm]  (6.5,7*\rr) -- (7.5,7*\rr)--(8,6*\rr)--(7.5,5*\rr);
\draw [line width=0.2mm]  (6.5,9*\rr) -- (7.5,9*\rr)--(8,8*\rr)--(7.5,7*\rr);
\draw [line width=0.2mm]  (6.5,11*\rr) -- (7.5,11*\rr)--(8,10*\rr)--(7.5,9*\rr);
\draw [line width=0.2mm]  (6.5,13*\rr) -- (7.5,13*\rr)--(8,12*\rr)--(7.5,11*\rr);
\draw [line width=0.2mm]  (6.5,15*\rr) -- (7.5,15*\rr)--(8,14*\rr)--(7.5,13*\rr);
\draw [line width=0.2mm]  (6.5,17*\rr) -- (7.5,17*\rr)--(8,16*\rr)--(7.5,15*\rr);
\draw [line width=0.2mm]  (6.5,19*\rr) -- (7.5,19*\rr)--(8,18*\rr)--(7.5,17*\rr);
\draw [line width=0.2mm]  (6.5,21*\rr) -- (7.5,21*\rr)--(8,20*\rr)--(7.5,19*\rr);

\draw [line width=0.2mm]  (8,0)--(9,0)--(9.5,-1*\rr);
\draw [line width=0.2mm]  (8,2*\rr)--(9,2*\rr)--(9.5,1*\rr)--(9,0);
\draw [line width=0.2mm]  (8,4*\rr)--(9,4*\rr)--(9.5,3*\rr)--(9,2*\rr);
\draw [line width=0.2mm]  (8,6*\rr)--(9,6*\rr)--(9.5,5*\rr)--(9,4*\rr);
\draw [line width=0.2mm]  (8,8*\rr)--(9,8*\rr)--(9.5,7*\rr)--(9,6*\rr);
\draw [line width=0.2mm]  (8,10*\rr)--(9,10*\rr)--(9.5,9*\rr)--(9,8*\rr);
\draw [line width=0.2mm]  (8,12*\rr)--(9,12*\rr)--(9.5,11*\rr)--(9,10*\rr);
\draw [line width=0.2mm]  (8,14*\rr)--(9,14*\rr)--(9.5,13*\rr)--(9,12*\rr);
\draw [line width=0.2mm]  (8,16*\rr)--(9,16*\rr)--(9.5,15*\rr)--(9,14*\rr);
\draw [line width=0.2mm]  (8,18*\rr)--(9,18*\rr)--(9.5,17*\rr)--(9,16*\rr);
\draw [line width=0.2mm]  (8,20*\rr)--(9,20*\rr)--(9.5,19*\rr)--(9,18*\rr);
\draw [line width=0.2mm]  (9.5,21*\rr)--(9,20*\rr);

\draw [line width=0.2mm]  (9.5,1*\rr)--(10.5,1*\rr)--(11,0)--(10.5,-1*\rr)
--(9.5,-1*\rr);
\draw [line width=0.2mm]  (9.5,3*\rr)--(10.5,3*\rr)--(11,2*\rr)--(10.5,1*\rr);
\draw [line width=0.2mm]  (9.5,5*\rr)--(10.5,5*\rr)--(11,4*\rr)--(10.5,3*\rr);
\draw [line width=0.2mm]  (9.5,7*\rr)--(10.5,7*\rr)--(11,6*\rr)--(10.5,5*\rr);
\draw [line width=0.2mm]  (9.5,9*\rr)--(10.5,9*\rr)--(11,8*\rr)--(10.5,7*\rr);
\draw [line width=0.2mm]  (9.5,11*\rr)--(10.5,11*\rr)--(11,10*\rr)--(10.5,9*\rr);
\draw [line width=0.2mm]  (9.5,13*\rr)--(10.5,13*\rr)--(11,12*\rr)--(10.5,11*\rr);
\draw [line width=0.2mm]  (9.5,15*\rr)--(10.5,15*\rr)--(11,14*\rr)--(10.5,13*\rr);
\draw [line width=0.2mm]  (9.5,17*\rr)--(10.5,17*\rr)--(11,16*\rr)--(10.5,15*\rr);
\draw [line width=0.2mm]  (9.5,19*\rr)--(10.5,19*\rr)--(11,18*\rr)--(10.5,17*\rr);
\draw [line width=0.2mm]  (9.5,21*\rr)--(10.5,21*\rr)--(11,20*\rr)--(10.5,19*\rr);

\draw [line width=0.2mm]  (11,0)--(12,0)--(12.5,-1*\rr);
\draw [line width=0.2mm]  (11,2*\rr)--(12,2*\rr)--(12.5,1*\rr)--(12,0);
\draw [line width=0.2mm]  (11,4*\rr)--(12,4*\rr)--(12.5,3*\rr)--(12,2*\rr);
\draw [line width=0.2mm]  (11,6*\rr)--(12,6*\rr)--(12.5,5*\rr)--(12,4*\rr);
\draw [line width=0.2mm]  (11,8*\rr)--(12,8*\rr)--(12.5,7*\rr)--(12,6*\rr);
\draw [line width=0.2mm]  (11,10*\rr)--(12,10*\rr)--(12.5,9*\rr)--(12,8*\rr);
\draw [line width=0.2mm]  (11,12*\rr)--(12,12*\rr)--(12.5,11*\rr)--(12,10*\rr);
\draw [line width=0.2mm]  (11,14*\rr)--(12,14*\rr)--(12.5,13*\rr)--(12,12*\rr);
\draw [line width=0.2mm]  (11,16*\rr)--(12,16*\rr)--(12.5,15*\rr)--(12,14*\rr);
\draw [line width=0.2mm]  (11,18*\rr)--(12,18*\rr)--(12.5,17*\rr)--(12,16*\rr);
\draw [line width=0.2mm]  (11,20*\rr)--(12,20*\rr)--(12.5,19*\rr)--(12,18*\rr);
\draw [line width=0.2mm]  (12,20*\rr)--(12.5,21*\rr);

\draw [line width=0.2mm]  (12.5,1*\rr)--(13.5,1*\rr)--(14,0)--(13.5,-1*\rr)
--(12.5,-1*\rr);
\draw [line width=0.2mm]  (12.5,3*\rr)--(13.5,3*\rr)--(14,2*\rr)--(13.5,1*\rr);
\draw [line width=0.2mm]  (12.5,5*\rr)--(13.5,5*\rr)--(14,4*\rr)--(13.5,3*\rr);
\draw [line width=0.2mm]  (12.5,7*\rr)--(13.5,7*\rr)--(14,6*\rr)--(13.5,5*\rr);
\draw [line width=0.2mm]  (12.5,9*\rr)--(13.5,9*\rr)--(14,8*\rr)--(13.5,7*\rr);
\draw [line width=0.2mm]  (12.5,11*\rr)--(13.5,11*\rr)--(14,10*\rr)--(13.5,9*\rr);
\draw [line width=0.2mm]  (12.5,13*\rr)--(13.5,13*\rr)--(14,12*\rr)--(13.5,11*\rr);
\draw [line width=0.2mm]  (12.5,15*\rr)--(13.5,15*\rr)--(14,14*\rr)--(13.5,13*\rr);
\draw [line width=0.2mm]  (12.5,17*\rr)--(13.5,17*\rr)--(14,16*\rr)--(13.5,15*\rr);
\draw [line width=0.2mm]  (12.5,19*\rr)--(13.5,19*\rr)--(14,18*\rr)--(13.5,17*\rr);
\draw [line width=0.2mm]  (12.5,21*\rr)--(13.5,21*\rr)--(14,20*\rr)--(13.5,19*\rr);

\draw [line width=0.2mm]  (14,0)--(15,0)--(15.5,-1*\rr);
\draw [line width=0.2mm]  (14,2*\rr)--(15,2*\rr)--(15.5,1*\rr)--(15,0);
\draw [line width=0.2mm]  (14,4*\rr)--(15,4*\rr)--(15.5,3*\rr)--(15,2*\rr);
\draw [line width=0.2mm]  (14,6*\rr)--(15,6*\rr)--(15.5,5*\rr)--(15,4*\rr);
\draw [line width=0.2mm]  (14,8*\rr)--(15,8*\rr)--(15.5,7*\rr)--(15,6*\rr);
\draw [line width=0.2mm]  (14,10*\rr)--(15,10*\rr)--(15.5,9*\rr)--(15,8*\rr);
\draw [line width=0.2mm]  (14,12*\rr)--(15,12*\rr)--(15.5,11*\rr)--(15,10*\rr);
\draw [line width=0.2mm]  (14,14*\rr)--(15,14*\rr)--(15.5,13*\rr)--(15,12*\rr);
\draw [line width=0.2mm]  (14,16*\rr)--(15,16*\rr)--(15.5,15*\rr)--(15,14*\rr);
\draw [line width=0.2mm]  (14,18*\rr)--(15,18*\rr)--(15.5,17*\rr)--(15,16*\rr);
\draw [line width=0.2mm]  (14,20*\rr)--(15,20*\rr)--(15.5,19*\rr)--(15,18*\rr);
\draw [line width=0.2mm]  (15,20*\rr)--(15.5,21*\rr);

\draw [line width=0.2mm]  (15.5,1*\rr)--(16.5,1*\rr)--(17,0)--(16.5,-1*\rr)
--(15.5,-1*\rr);
\draw [line width=0.2mm]  (15.5,3*\rr)--(16.5,3*\rr)--(17,2*\rr)--(16.5,1*\rr);
\draw [line width=0.2mm]  (15.5,5*\rr)--(16.5,5*\rr)--(17,4*\rr)--(16.5,3*\rr);
\draw [line width=0.2mm]  (15.5,7*\rr)--(16.5,7*\rr)--(17,6*\rr)--(16.5,5*\rr);
\draw [line width=0.2mm]  (15.5,9*\rr)--(16.5,9*\rr)--(17,8*\rr)--(16.5,7*\rr);
\draw [line width=0.2mm]  (15.5,11*\rr)--(16.5,11*\rr)--(17,10*\rr)--(16.5,9*\rr);
\draw [line width=0.2mm]  (15.5,13*\rr)--(16.5,13*\rr)--(17,12*\rr)--(16.5,11*\rr);
\draw [line width=0.2mm]  (15.5,15*\rr)--(16.5,15*\rr)--(17,14*\rr)--(16.5,13*\rr);
\draw [line width=0.2mm]  (15.5,17*\rr)--(16.5,17*\rr)--(17,16*\rr)--(16.5,15*\rr);
\draw [line width=0.2mm]  (15.5,19*\rr)--(16.5,19*\rr)--(17,18*\rr)--(16.5,17*\rr);
\draw [line width=0.2mm]  (15.5,21*\rr)--(16.5,21*\rr)--(17,20*\rr)--(16.5,19*\rr);

\draw [line width=0.2mm]  (17,0)--(18,0)--(18.5,-1*\rr);
\draw [line width=0.2mm]  (17,2*\rr)--(18,2*\rr)--(18.5,1*\rr)--(18,0);
\draw [line width=0.2mm]  (17,4*\rr)--(18,4*\rr)--(18.5,3*\rr)--(18,2*\rr);
\draw [line width=0.2mm]  (17,6*\rr)--(18,6*\rr)--(18.5,5*\rr)--(18,4*\rr);
\draw [line width=0.2mm]  (17,8*\rr)--(18,8*\rr)--(18.5,7*\rr)--(18,6*\rr);
\draw [line width=0.2mm]  (17,10*\rr)--(18,10*\rr)--(18.5,9*\rr)--(18,8*\rr);
\draw [line width=0.2mm]  (17,12*\rr)--(18,12*\rr)--(18.5,11*\rr)--(18,10*\rr);
\draw [line width=0.2mm]  (17,14*\rr)--(18,14*\rr)--(18.5,13*\rr)--(18,12*\rr);
\draw [line width=0.2mm]  (17,16*\rr)--(18,16*\rr)--(18.5,15*\rr)--(18,14*\rr);
\draw [line width=0.2mm]  (17,18*\rr)--(18,18*\rr)--(18.5,17*\rr)--(18,16*\rr);
\draw [line width=0.2mm]  (17,20*\rr)--(18,20*\rr)--(18.5,19*\rr)--(18,18*\rr);
\draw [line width=0.2mm]  (18,20*\rr)--(18.5,21*\rr);

\draw [line width=0.2mm]  (18.5,1*\rr)--(19.5,1*\rr)--(20,0)--(19.5,-1*\rr)
--(18.5,-1*\rr);
\draw [line width=0.2mm]  (18.5,3*\rr)--(19.5,3*\rr)--(20,2*\rr)--(19.5,1*\rr);
\draw [line width=0.2mm]  (18.5,5*\rr)--(19.5,5*\rr)--(20,4*\rr)--(19.5,3*\rr);
\draw [line width=0.2mm]  (18.5,7*\rr)--(19.5,7*\rr)--(20,6*\rr)--(19.5,5*\rr);
\draw [line width=0.2mm]  (18.5,9*\rr)--(19.5,9*\rr)--(20,8*\rr)--(19.5,7*\rr);
\draw [line width=0.2mm]  (18.5,11*\rr)--(19.5,11*\rr)--(20,10*\rr)--(19.5,9*\rr);
\draw [line width=0.2mm]  (18.5,13*\rr)--(19.5,13*\rr)--(20,12*\rr)--(19.5,11*\rr);
\draw [line width=0.2mm]  (18.5,15*\rr)--(19.5,15*\rr)--(20,14*\rr)--(19.5,13*\rr);
\draw [line width=0.2mm]  (18.5,17*\rr)--(19.5,17*\rr)--(20,16*\rr)--(19.5,15*\rr);
\draw [line width=0.2mm]  (18.5,19*\rr)--(19.5,19*\rr)--(20,18*\rr)--(19.5,17*\rr);
\draw [line width=0.2mm]  (18.5,21*\rr)--(19.5,21*\rr)--(20,20*\rr)--(19.5,19*\rr);

\draw [line width=0.2mm]  (20,0)--(21,0)--(21.5,-1*\rr);
\draw [line width=0.2mm]  (20,2*\rr)--(21,2*\rr)--(21.5,1*\rr)--(21,0);
\draw [line width=0.2mm]  (20,4*\rr)--(21,4*\rr)--(21.5,3*\rr)--(21,2*\rr);
\draw [line width=0.2mm]  (20,6*\rr)--(21,6*\rr)--(21.5,5*\rr)--(21,4*\rr);
\draw [line width=0.2mm]  (20,8*\rr)--(21,8*\rr)--(21.5,7*\rr)--(21,6*\rr);
\draw [line width=0.2mm]  (20,10*\rr)--(21,10*\rr)--(21.5,9*\rr)--(21,8*\rr);
\draw [line width=0.2mm]  (20,12*\rr)--(21,12*\rr)--(21.5,11*\rr)--(21,10*\rr);
\draw [line width=0.2mm]  (20,14*\rr)--(21,14*\rr)--(21.5,13*\rr)--(21,12*\rr);
\draw [line width=0.2mm]  (20,16*\rr)--(21,16*\rr)--(21.5,15*\rr)--(21,14*\rr);
\draw [line width=0.2mm]  (20,18*\rr)--(21,18*\rr)--(21.5,17*\rr)--(21,16*\rr);
\draw [line width=0.2mm]  (20,20*\rr)--(21,20*\rr)--(21.5,19*\rr)--(21,18*\rr);
\draw [line width=0.2mm]  (21,20*\rr)--(21.5,21*\rr);

\draw [line width=0.2mm]  (21.5,1*\rr)--(22.5,1*\rr)--(23,0)--(22.5,-1*\rr)
--(21.5,-1*\rr);
\draw [line width=0.2mm]  (21.5,3*\rr)--(22.5,3*\rr)--(23,2*\rr)--(22.5,1*\rr);
\draw [line width=0.2mm]  (21.5,5*\rr)--(22.5,5*\rr)--(23,4*\rr)--(22.5,3*\rr);
\draw [line width=0.2mm]  (21.5,7*\rr)--(22.5,7*\rr)--(23,6*\rr)--(22.5,5*\rr);
\draw [line width=0.2mm]  (21.5,9*\rr)--(22.5,9*\rr)--(23,8*\rr)--(22.5,7*\rr);
\draw [line width=0.2mm]  (21.5,11*\rr)--(22.5,11*\rr)--(23,10*\rr)--(22.5,9*\rr);
\draw [line width=0.2mm]  (21.5,13*\rr)--(22.5,13*\rr)--(23,12*\rr)--(22.5,11*\rr);
\draw [line width=0.2mm]  (21.5,15*\rr)--(22.5,15*\rr)--(23,14*\rr)--(22.5,13*\rr);
\draw [line width=0.2mm]  (21.5,17*\rr)--(22.5,17*\rr)--(23,16*\rr)--(22.5,15*\rr);
\draw [line width=0.2mm]  (21.5,19*\rr)--(22.5,19*\rr)--(23,18*\rr)--(22.5,17*\rr);
\draw [line width=0.2mm]  (21.5,21*\rr)--(22.5,21*\rr)--(23,20*\rr)--(22.5,19*\rr);

\draw [line width=0.2mm]  (23,0)--(24,0)--(24.5,-1*\rr);
\draw [line width=0.2mm]  (23,2*\rr)--(24,2*\rr)--(24.5,1*\rr)--(24,0);
\draw [line width=0.2mm]  (23,4*\rr)--(24,4*\rr)--(24.5,3*\rr)--(24,2*\rr);
\draw [line width=0.2mm]  (23,6*\rr)--(24,6*\rr)--(24.5,5*\rr)--(24,4*\rr);
\draw [line width=0.2mm]  (23,8*\rr)--(24,8*\rr)--(24.5,7*\rr)--(24,6*\rr);
\draw [line width=0.2mm]  (23,10*\rr)--(24,10*\rr)--(24.5,9*\rr)--(24,8*\rr);
\draw [line width=0.2mm]  (23,12*\rr)--(24,12*\rr)--(24.5,11*\rr)--(24,10*\rr);
\draw [line width=0.2mm]  (23,14*\rr)--(24,14*\rr)--(24.5,13*\rr)--(24,12*\rr);
\draw [line width=0.2mm]  (23,16*\rr)--(24,16*\rr)--(24.5,15*\rr)--(24,14*\rr);
\draw [line width=0.2mm]  (23,18*\rr)--(24,18*\rr)--(24.5,17*\rr)--(24,16*\rr);
\draw [line width=0.2mm]  (23,20*\rr)--(24,20*\rr)--(24.5,19*\rr)--(24,18*\rr);
\draw [line width=0.2mm]  (24,20*\rr)--(24.5,21*\rr);

\draw [line width=0.5mm] (2,0)--(2,8*\rr)--(8,4*\rr)--(2,0);
\draw [line width=0.5mm] (2,8*\rr)--(8,12*\rr)--(8,4*\rr);
\draw [line width=0.5mm] (8,12*\rr)--(14,8*\rr)--(8,4*\rr)--(14,0)--(14,8*\rr);

\foreach \pos in {(2,0), (2,8*\rr),(8,12*\rr),(8,4*\rr),(14,0),(14,8*\rr)}
\shade[shading=ball, ball color=lightgray] \pos circle (.6);
 
\end{tikzpicture}\qquad (b) \begin{tikzpicture}[scale=0.29]
\clip (-1, -1.6*\rr) rectangle (20, 13.5*\rr);

\draw [line width=0.2mm] (0.5,1*\rr) -- (1.5,1*\rr) -- (2,0) -- (1.5,-1*\rr)
-- (0.5,-1*\rr) -- (0,0) -- (0.5,1*\rr);
\draw [line width=0.2mm] (0.5,1*\rr) -- (0,2*\rr) -- (0.5,3*\rr) -- (1.5,3*\rr) 
-- (2,2*\rr) -- (1.5,1*\rr);
\draw [line width=0.2mm] (0.5,3*\rr) -- (0,4*\rr) -- (0.5,5*\rr) -- (1.5,5*\rr) 
-- (2,4*\rr) -- (1.5,3*\rr);
\draw [line width=0.2mm] (0.5,5*\rr) -- (0,6*\rr) -- (0.5,7*\rr) -- (1.5,7*\rr) 
-- (2,6*\rr) -- (1.5,5*\rr);
\draw [line width=0.2mm] (0.5,7*\rr) -- (0,8*\rr) -- (0.5,9*\rr) -- (1.5,9*\rr) 
-- (2,8*\rr) -- (1.5,7*\rr);
\draw [line width=0.2mm] (0.5,9*\rr) -- (0,10*\rr) -- (0.5,11*\rr) -- (1.5,11*\rr) 
-- (2,10*\rr) -- (1.5,9*\rr);
\draw [line width=0.2mm] (0.5,11*\rr) -- (0,12*\rr) -- (0.5,13*\rr) -- (1.5,13*\rr) 
-- (2,12*\rr) -- (1.5,11*\rr);
\draw [line width=0.2mm] (0.5,13*\rr) -- (0,14*\rr) -- (0.5,15*\rr) -- (1.5,15*\rr) 
-- (2,14*\rr) -- (1.5,13*\rr);
\draw [line width=0.2mm] (0.5,15*\rr) -- (0,16*\rr) -- (0.5,17*\rr) -- (1.5,17*\rr) 
-- (2,16*\rr) -- (1.5,15*\rr);
\draw [line width=0.2mm] (0.5,17*\rr) -- (0,18*\rr) -- (0.5,19*\rr) -- (1.5,19*\rr) 
-- (2,18*\rr) -- (1.5,17*\rr);
\draw [line width=0.2mm] (0.5,19*\rr) -- (0,20*\rr) -- (0.5,21*\rr) -- (1.5,21*\rr) 
-- (2,20*\rr) -- (1.5,19*\rr);

\draw [line width=0.2mm]  (3,0) -- (3.5,-1*\rr);
\draw [line width=0.2mm] (2,2*\rr) -- (3,2*\rr) -- (3.5,1*\rr) -- (3,0) -- (2,0);
\draw [line width=0.2mm] (2,4*\rr) -- (3,4*\rr) -- (3.5,3*\rr) -- (3,2*\rr);
\draw [line width=0.2mm] (2,6*\rr) -- (3,6*\rr) -- (3.5,5*\rr) -- (3,4*\rr);
\draw [line width=0.2mm] (2,8*\rr) -- (3,8*\rr) -- (3.5,7*\rr) -- (3,6*\rr);
\draw [line width=0.2mm] (2,10*\rr) -- (3,10*\rr) -- (3.5,9*\rr) -- (3,8*\rr);
\draw [line width=0.2mm] (2,12*\rr) -- (3,12*\rr) -- (3.5,11*\rr) -- (3,10*\rr);
\draw [line width=0.2mm] (2,14*\rr) -- (3,14*\rr) -- (3.5,13*\rr) -- (3,12*\rr);
\draw [line width=0.2mm] (2,16*\rr) -- (3,16*\rr) -- (3.5,15*\rr) -- (3,14*\rr);
\draw [line width=0.2mm] (2,18*\rr) -- (3,18*\rr) -- (3.5,17*\rr) -- (3,16*\rr);
\draw [line width=0.2mm] (2,20*\rr) -- (3,20*\rr) -- (3.5,19*\rr) -- (3,18*\rr);
\draw [line width=0.2mm] (3.5,21*\rr) -- (3,20*\rr);

\draw [line width=0.2mm]  (3.5,1*\rr) -- (4.5,1*\rr) -- (5,0) -- (4.5,-1*\rr)
--(3.5,-1*\rr);
\draw [line width=0.2mm]  (3.5,3*\rr) -- (4.5,3*\rr) -- (5,2*\rr) -- (4.5,1*\rr);
\draw [line width=0.2mm]  (3.5,5*\rr) -- (4.5,5*\rr) -- (5,4*\rr) -- (4.5,3*\rr);
\draw [line width=0.2mm]  (3.5,7*\rr) -- (4.5,7*\rr) -- (5,6*\rr) -- (4.5,5*\rr);
\draw [line width=0.2mm]  (3.5,9*\rr) -- (4.5,9*\rr) -- (5,8*\rr) -- (4.5,7*\rr);
\draw [line width=0.2mm]  (3.5,11*\rr) -- (4.5,11*\rr) -- (5,10*\rr) -- (4.5,9*\rr);
\draw [line width=0.2mm]  (3.5,13*\rr) -- (4.5,13*\rr) -- (5,12*\rr) -- (4.5,11*\rr);
\draw [line width=0.2mm]  (3.5,15*\rr) -- (4.5,15*\rr) -- (5,14*\rr) -- (4.5,13*\rr);
\draw [line width=0.2mm]  (3.5,17*\rr) -- (4.5,17*\rr) -- (5,16*\rr) -- (4.5,15*\rr);
\draw [line width=0.2mm]  (3.5,19*\rr) -- (4.5,19*\rr) -- (5,18*\rr) -- (4.5,17*\rr);
\draw [line width=0.2mm]  (3.5,21*\rr) -- (4.5,21*\rr) -- (5,20*\rr) -- (4.5,19*\rr);

\draw [line width=0.2mm]  (5,0) -- (6,0)--(6.5,-1*\rr );
\draw [line width=0.2mm]  (5,2*\rr) -- (6,2*\rr)--(6.5,1*\rr )--(6,0);
\draw [line width=0.2mm]  (5,4*\rr) -- (6,4*\rr)--(6.5,3*\rr )--(6,2*\rr);
\draw [line width=0.2mm]  (5,6*\rr) -- (6,6*\rr)--(6.5,5*\rr )--(6,4*\rr);
\draw [line width=0.2mm]  (5,8*\rr) -- (6,8*\rr)--(6.5,7*\rr )--(6,6*\rr);
\draw [line width=0.2mm]  (5,10*\rr) -- (6,10*\rr)--(6.5,9*\rr )--(6,8*\rr);
\draw [line width=0.2mm]  (5,12*\rr) -- (6,12*\rr)--(6.5,11*\rr )--(6,10*\rr);
\draw [line width=0.2mm]  (5,14*\rr) -- (6,14*\rr)--(6.5,13*\rr )--(6,12*\rr);
\draw [line width=0.2mm]  (5,16*\rr) -- (6,16*\rr)--(6.5,15*\rr )--(6,14*\rr);
\draw [line width=0.2mm]  (5,18*\rr) -- (6,18*\rr)--(6.5,17*\rr )--(6,16*\rr);
\draw [line width=0.2mm]  (5,20*\rr) -- (6,20*\rr)--(6.5,19*\rr )--(6,18*\rr);
\draw [line width=0.2mm]  (6.5,21*\rr )--(6,20*\rr);

\draw [line width=0.2mm]  (6.5,1*\rr) -- (7.5,1*\rr)--(8,0)--(7.5,-1*\rr)
--(6.5,-1*\rr);
\draw [line width=0.2mm]  (6.5,3*\rr) -- (7.5,3*\rr)--(8,2*\rr)--(7.5,1*\rr);
\draw [line width=0.2mm]  (6.5,5*\rr) -- (7.5,5*\rr)--(8,4*\rr)--(7.5,3*\rr);
\draw [line width=0.2mm]  (6.5,7*\rr) -- (7.5,7*\rr)--(8,6*\rr)--(7.5,5*\rr);
\draw [line width=0.2mm]  (6.5,9*\rr) -- (7.5,9*\rr)--(8,8*\rr)--(7.5,7*\rr);
\draw [line width=0.2mm]  (6.5,11*\rr) -- (7.5,11*\rr)--(8,10*\rr)--(7.5,9*\rr);
\draw [line width=0.2mm]  (6.5,13*\rr) -- (7.5,13*\rr)--(8,12*\rr)--(7.5,11*\rr);
\draw [line width=0.2mm]  (6.5,15*\rr) -- (7.5,15*\rr)--(8,14*\rr)--(7.5,13*\rr);
\draw [line width=0.2mm]  (6.5,17*\rr) -- (7.5,17*\rr)--(8,16*\rr)--(7.5,15*\rr);
\draw [line width=0.2mm]  (6.5,19*\rr) -- (7.5,19*\rr)--(8,18*\rr)--(7.5,17*\rr);
\draw [line width=0.2mm]  (6.5,21*\rr) -- (7.5,21*\rr)--(8,20*\rr)--(7.5,19*\rr);

\draw [line width=0.2mm]  (8,0)--(9,0)--(9.5,-1*\rr);
\draw [line width=0.2mm]  (8,2*\rr)--(9,2*\rr)--(9.5,1*\rr)--(9,0);
\draw [line width=0.2mm]  (8,4*\rr)--(9,4*\rr)--(9.5,3*\rr)--(9,2*\rr);
\draw [line width=0.2mm]  (8,6*\rr)--(9,6*\rr)--(9.5,5*\rr)--(9,4*\rr);
\draw [line width=0.2mm]  (8,8*\rr)--(9,8*\rr)--(9.5,7*\rr)--(9,6*\rr);
\draw [line width=0.2mm]  (8,10*\rr)--(9,10*\rr)--(9.5,9*\rr)--(9,8*\rr);
\draw [line width=0.2mm]  (8,12*\rr)--(9,12*\rr)--(9.5,11*\rr)--(9,10*\rr);
\draw [line width=0.2mm]  (8,14*\rr)--(9,14*\rr)--(9.5,13*\rr)--(9,12*\rr);
\draw [line width=0.2mm]  (8,16*\rr)--(9,16*\rr)--(9.5,15*\rr)--(9,14*\rr);
\draw [line width=0.2mm]  (8,18*\rr)--(9,18*\rr)--(9.5,17*\rr)--(9,16*\rr);
\draw [line width=0.2mm]  (8,20*\rr)--(9,20*\rr)--(9.5,19*\rr)--(9,18*\rr);
\draw [line width=0.2mm]  (9.5,21*\rr)--(9,20*\rr);

\draw [line width=0.2mm]  (9.5,1*\rr)--(10.5,1*\rr)--(11,0)--(10.5,-1*\rr)
--(9.5,-1*\rr);
\draw [line width=0.2mm]  (9.5,3*\rr)--(10.5,3*\rr)--(11,2*\rr)--(10.5,1*\rr);
\draw [line width=0.2mm]  (9.5,5*\rr)--(10.5,5*\rr)--(11,4*\rr)--(10.5,3*\rr);
\draw [line width=0.2mm]  (9.5,7*\rr)--(10.5,7*\rr)--(11,6*\rr)--(10.5,5*\rr);
\draw [line width=0.2mm]  (9.5,9*\rr)--(10.5,9*\rr)--(11,8*\rr)--(10.5,7*\rr);
\draw [line width=0.2mm]  (9.5,11*\rr)--(10.5,11*\rr)--(11,10*\rr)--(10.5,9*\rr);
\draw [line width=0.2mm]  (9.5,13*\rr)--(10.5,13*\rr)--(11,12*\rr)--(10.5,11*\rr);
\draw [line width=0.2mm]  (9.5,15*\rr)--(10.5,15*\rr)--(11,14*\rr)--(10.5,13*\rr);
\draw [line width=0.2mm]  (9.5,17*\rr)--(10.5,17*\rr)--(11,16*\rr)--(10.5,15*\rr);
\draw [line width=0.2mm]  (9.5,19*\rr)--(10.5,19*\rr)--(11,18*\rr)--(10.5,17*\rr);
\draw [line width=0.2mm]  (9.5,21*\rr)--(10.5,21*\rr)--(11,20*\rr)--(10.5,19*\rr);

\draw [line width=0.2mm]  (11,0)--(12,0)--(12.5,-1*\rr);
\draw [line width=0.2mm]  (11,2*\rr)--(12,2*\rr)--(12.5,1*\rr)--(12,0);
\draw [line width=0.2mm]  (11,4*\rr)--(12,4*\rr)--(12.5,3*\rr)--(12,2*\rr);
\draw [line width=0.2mm]  (11,6*\rr)--(12,6*\rr)--(12.5,5*\rr)--(12,4*\rr);
\draw [line width=0.2mm]  (11,8*\rr)--(12,8*\rr)--(12.5,7*\rr)--(12,6*\rr);
\draw [line width=0.2mm]  (11,10*\rr)--(12,10*\rr)--(12.5,9*\rr)--(12,8*\rr);
\draw [line width=0.2mm]  (11,12*\rr)--(12,12*\rr)--(12.5,11*\rr)--(12,10*\rr);
\draw [line width=0.2mm]  (11,14*\rr)--(12,14*\rr)--(12.5,13*\rr)--(12,12*\rr);
\draw [line width=0.2mm]  (11,16*\rr)--(12,16*\rr)--(12.5,15*\rr)--(12,14*\rr);
\draw [line width=0.2mm]  (11,18*\rr)--(12,18*\rr)--(12.5,17*\rr)--(12,16*\rr);
\draw [line width=0.2mm]  (11,20*\rr)--(12,20*\rr)--(12.5,19*\rr)--(12,18*\rr);
\draw [line width=0.2mm]  (12,20*\rr)--(12.5,21*\rr);

\draw [line width=0.2mm]  (12.5,1*\rr)--(13.5,1*\rr)--(14,0)--(13.5,-1*\rr)
--(12.5,-1*\rr);
\draw [line width=0.2mm]  (12.5,3*\rr)--(13.5,3*\rr)--(14,2*\rr)--(13.5,1*\rr);
\draw [line width=0.2mm]  (12.5,5*\rr)--(13.5,5*\rr)--(14,4*\rr)--(13.5,3*\rr);
\draw [line width=0.2mm]  (12.5,7*\rr)--(13.5,7*\rr)--(14,6*\rr)--(13.5,5*\rr);
\draw [line width=0.2mm]  (12.5,9*\rr)--(13.5,9*\rr)--(14,8*\rr)--(13.5,7*\rr);
\draw [line width=0.2mm]  (12.5,11*\rr)--(13.5,11*\rr)--(14,10*\rr)--(13.5,9*\rr);
\draw [line width=0.2mm]  (12.5,13*\rr)--(13.5,13*\rr)--(14,12*\rr)--(13.5,11*\rr);
\draw [line width=0.2mm]  (12.5,15*\rr)--(13.5,15*\rr)--(14,14*\rr)--(13.5,13*\rr);
\draw [line width=0.2mm]  (12.5,17*\rr)--(13.5,17*\rr)--(14,16*\rr)--(13.5,15*\rr);
\draw [line width=0.2mm]  (12.5,19*\rr)--(13.5,19*\rr)--(14,18*\rr)--(13.5,17*\rr);
\draw [line width=0.2mm]  (12.5,21*\rr)--(13.5,21*\rr)--(14,20*\rr)--(13.5,19*\rr);

\draw [line width=0.2mm]  (14,0)--(15,0)--(15.5,-1*\rr);
\draw [line width=0.2mm]  (14,2*\rr)--(15,2*\rr)--(15.5,1*\rr)--(15,0);
\draw [line width=0.2mm]  (14,4*\rr)--(15,4*\rr)--(15.5,3*\rr)--(15,2*\rr);
\draw [line width=0.2mm]  (14,6*\rr)--(15,6*\rr)--(15.5,5*\rr)--(15,4*\rr);
\draw [line width=0.2mm]  (14,8*\rr)--(15,8*\rr)--(15.5,7*\rr)--(15,6*\rr);
\draw [line width=0.2mm]  (14,10*\rr)--(15,10*\rr)--(15.5,9*\rr)--(15,8*\rr);
\draw [line width=0.2mm]  (14,12*\rr)--(15,12*\rr)--(15.5,11*\rr)--(15,10*\rr);
\draw [line width=0.2mm]  (14,14*\rr)--(15,14*\rr)--(15.5,13*\rr)--(15,12*\rr);
\draw [line width=0.2mm]  (14,16*\rr)--(15,16*\rr)--(15.5,15*\rr)--(15,14*\rr);
\draw [line width=0.2mm]  (14,18*\rr)--(15,18*\rr)--(15.5,17*\rr)--(15,16*\rr);
\draw [line width=0.2mm]  (14,20*\rr)--(15,20*\rr)--(15.5,19*\rr)--(15,18*\rr);
\draw [line width=0.2mm]  (15,20*\rr)--(15.5,21*\rr);

\draw [line width=0.2mm]  (15.5,1*\rr)--(16.5,1*\rr)--(17,0)--(16.5,-1*\rr)
--(15.5,-1*\rr);
\draw [line width=0.2mm]  (15.5,3*\rr)--(16.5,3*\rr)--(17,2*\rr)--(16.5,1*\rr);
\draw [line width=0.2mm]  (15.5,5*\rr)--(16.5,5*\rr)--(17,4*\rr)--(16.5,3*\rr);
\draw [line width=0.2mm]  (15.5,7*\rr)--(16.5,7*\rr)--(17,6*\rr)--(16.5,5*\rr);
\draw [line width=0.2mm]  (15.5,9*\rr)--(16.5,9*\rr)--(17,8*\rr)--(16.5,7*\rr);
\draw [line width=0.2mm]  (15.5,11*\rr)--(16.5,11*\rr)--(17,10*\rr)--(16.5,9*\rr);
\draw [line width=0.2mm]  (15.5,13*\rr)--(16.5,13*\rr)--(17,12*\rr)--(16.5,11*\rr);
\draw [line width=0.2mm]  (15.5,15*\rr)--(16.5,15*\rr)--(17,14*\rr)--(16.5,13*\rr);
\draw [line width=0.2mm]  (15.5,17*\rr)--(16.5,17*\rr)--(17,16*\rr)--(16.5,15*\rr);
\draw [line width=0.2mm]  (15.5,19*\rr)--(16.5,19*\rr)--(17,18*\rr)--(16.5,17*\rr);
\draw [line width=0.2mm]  (15.5,21*\rr)--(16.5,21*\rr)--(17,20*\rr)--(16.5,19*\rr);

\draw [line width=0.2mm]  (17,0)--(18,0)--(18.5,-1*\rr);
\draw [line width=0.2mm]  (17,2*\rr)--(18,2*\rr)--(18.5,1*\rr)--(18,0);
\draw [line width=0.2mm]  (17,4*\rr)--(18,4*\rr)--(18.5,3*\rr)--(18,2*\rr);
\draw [line width=0.2mm]  (17,6*\rr)--(18,6*\rr)--(18.5,5*\rr)--(18,4*\rr);
\draw [line width=0.2mm]  (17,8*\rr)--(18,8*\rr)--(18.5,7*\rr)--(18,6*\rr);
\draw [line width=0.2mm]  (17,10*\rr)--(18,10*\rr)--(18.5,9*\rr)--(18,8*\rr);
\draw [line width=0.2mm]  (17,12*\rr)--(18,12*\rr)--(18.5,11*\rr)--(18,10*\rr);
\draw [line width=0.2mm]  (17,14*\rr)--(18,14*\rr)--(18.5,13*\rr)--(18,12*\rr);
\draw [line width=0.2mm]  (17,16*\rr)--(18,16*\rr)--(18.5,15*\rr)--(18,14*\rr);
\draw [line width=0.2mm]  (17,18*\rr)--(18,18*\rr)--(18.5,17*\rr)--(18,16*\rr);
\draw [line width=0.2mm]  (17,20*\rr)--(18,20*\rr)--(18.5,19*\rr)--(18,18*\rr);
\draw [line width=0.2mm]  (18,20*\rr)--(18.5,21*\rr);

\draw [line width=0.2mm]  (18.5,1*\rr)--(19.5,1*\rr)--(20,0)--(19.5,-1*\rr)
--(18.5,-1*\rr);
\draw [line width=0.2mm]  (18.5,3*\rr)--(19.5,3*\rr)--(20,2*\rr)--(19.5,1*\rr);
\draw [line width=0.2mm]  (18.5,5*\rr)--(19.5,5*\rr)--(20,4*\rr)--(19.5,3*\rr);
\draw [line width=0.2mm]  (18.5,7*\rr)--(19.5,7*\rr)--(20,6*\rr)--(19.5,5*\rr);
\draw [line width=0.2mm]  (18.5,9*\rr)--(19.5,9*\rr)--(20,8*\rr)--(19.5,7*\rr);
\draw [line width=0.2mm]  (18.5,11*\rr)--(19.5,11*\rr)--(20,10*\rr)--(19.5,9*\rr);
\draw [line width=0.2mm]  (18.5,13*\rr)--(19.5,13*\rr)--(20,12*\rr)--(19.5,11*\rr);
\draw [line width=0.2mm]  (18.5,15*\rr)--(19.5,15*\rr)--(20,14*\rr)--(19.5,13*\rr);
\draw [line width=0.2mm]  (18.5,17*\rr)--(19.5,17*\rr)--(20,16*\rr)--(19.5,15*\rr);
\draw [line width=0.2mm]  (18.5,19*\rr)--(19.5,19*\rr)--(20,18*\rr)--(19.5,17*\rr);
\draw [line width=0.2mm]  (18.5,21*\rr)--(19.5,21*\rr)--(20,20*\rr)--(19.5,19*\rr);

\draw [line width=0.2mm]  (20,0)--(21,0)--(21.5,-1*\rr);
\draw [line width=0.2mm]  (20,2*\rr)--(21,2*\rr)--(21.5,1*\rr)--(21,0);
\draw [line width=0.2mm]  (20,4*\rr)--(21,4*\rr)--(21.5,3*\rr)--(21,2*\rr);
\draw [line width=0.2mm]  (20,6*\rr)--(21,6*\rr)--(21.5,5*\rr)--(21,4*\rr);
\draw [line width=0.2mm]  (20,8*\rr)--(21,8*\rr)--(21.5,7*\rr)--(21,6*\rr);
\draw [line width=0.2mm]  (20,10*\rr)--(21,10*\rr)--(21.5,9*\rr)--(21,8*\rr);
\draw [line width=0.2mm]  (20,12*\rr)--(21,12*\rr)--(21.5,11*\rr)--(21,10*\rr);
\draw [line width=0.2mm]  (20,14*\rr)--(21,14*\rr)--(21.5,13*\rr)--(21,12*\rr);
\draw [line width=0.2mm]  (20,16*\rr)--(21,16*\rr)--(21.5,15*\rr)--(21,14*\rr);
\draw [line width=0.2mm]  (20,18*\rr)--(21,18*\rr)--(21.5,17*\rr)--(21,16*\rr);
\draw [line width=0.2mm]  (20,20*\rr)--(21,20*\rr)--(21.5,19*\rr)--(21,18*\rr);
\draw [line width=0.2mm]  (21,20*\rr)--(21.5,21*\rr);

\draw [line width=0.2mm]  (21.5,1*\rr)--(22.5,1*\rr)--(23,0)--(22.5,-1*\rr)
--(21.5,-1*\rr);
\draw [line width=0.2mm]  (21.5,3*\rr)--(22.5,3*\rr)--(23,2*\rr)--(22.5,1*\rr);
\draw [line width=0.2mm]  (21.5,5*\rr)--(22.5,5*\rr)--(23,4*\rr)--(22.5,3*\rr);
\draw [line width=0.2mm]  (21.5,7*\rr)--(22.5,7*\rr)--(23,6*\rr)--(22.5,5*\rr);
\draw [line width=0.2mm]  (21.5,9*\rr)--(22.5,9*\rr)--(23,8*\rr)--(22.5,7*\rr);
\draw [line width=0.2mm]  (21.5,11*\rr)--(22.5,11*\rr)--(23,10*\rr)--(22.5,9*\rr);
\draw [line width=0.2mm]  (21.5,13*\rr)--(22.5,13*\rr)--(23,12*\rr)--(22.5,11*\rr);
\draw [line width=0.2mm]  (21.5,15*\rr)--(22.5,15*\rr)--(23,14*\rr)--(22.5,13*\rr);
\draw [line width=0.2mm]  (21.5,17*\rr)--(22.5,17*\rr)--(23,16*\rr)--(22.5,15*\rr);
\draw [line width=0.2mm]  (21.5,19*\rr)--(22.5,19*\rr)--(23,18*\rr)--(22.5,17*\rr);
\draw [line width=0.2mm]  (21.5,21*\rr)--(22.5,21*\rr)--(23,20*\rr)--(22.5,19*\rr);

\draw [line width=0.2mm]  (23,0)--(24,0)--(24.5,-1*\rr);
\draw [line width=0.2mm]  (23,2*\rr)--(24,2*\rr)--(24.5,1*\rr)--(24,0);
\draw [line width=0.2mm]  (23,4*\rr)--(24,4*\rr)--(24.5,3*\rr)--(24,2*\rr);
\draw [line width=0.2mm]  (23,6*\rr)--(24,6*\rr)--(24.5,5*\rr)--(24,4*\rr);
\draw [line width=0.2mm]  (23,8*\rr)--(24,8*\rr)--(24.5,7*\rr)--(24,6*\rr);
\draw [line width=0.2mm]  (23,10*\rr)--(24,10*\rr)--(24.5,9*\rr)--(24,8*\rr);
\draw [line width=0.2mm]  (23,12*\rr)--(24,12*\rr)--(24.5,11*\rr)--(24,10*\rr);
\draw [line width=0.2mm]  (23,14*\rr)--(24,14*\rr)--(24.5,13*\rr)--(24,12*\rr);
\draw [line width=0.2mm]  (23,16*\rr)--(24,16*\rr)--(24.5,15*\rr)--(24,14*\rr);
\draw [line width=0.2mm]  (23,18*\rr)--(24,18*\rr)--(24.5,17*\rr)--(24,16*\rr);
\draw [line width=0.2mm]  (23,20*\rr)--(24,20*\rr)--(24.5,19*\rr)--(24,18*\rr);
\draw [line width=0.2mm]  (24,20*\rr)--(24.5,21*\rr);

\draw [line width=0.5mm] (6,2*\rr)--(0,0)--(1.5,7*\rr)--(6,2*\rr)--(7.5,9*\rr)
--(1.5,7*\rr);
\draw [line width=0.5mm] (6,2*\rr)--(12,4*\rr)--(7.5,9*\rr)--(13.5,11*\rr)
--(12,4*\rr);
\draw [line width=0.5mm] (13.5,11*\rr)--(18,6*\rr)--(12,4*\rr)
--(16.5,-1*\rr)--(18,6*\rr);
\draw [line width=0.5mm] (13.5,11*\rr)--(19.5,13*\rr)--(18,6*\rr);

\foreach \pos in {(0,0),(1.5,7*\rr),(6,2*\rr),(7.5,9*\rr),(12,4*\rr),(13.5,11*\rr),
(16.5,-1*\rr),(18,6*\rr),(19.5,13*\rr)}
\shade[shading=ball, ball color=lightgray] \pos circle (.6);
 \end{tikzpicture} 
\caption{PGSs on ${\mathbb H}_2$, for $D^2=$ 48 (Case HA1) (a), and $D^2=$ 39 
(Case HA2) (b). The number of PGSs is 32 and 52, respectively.}
\end{center}
\end{figure}
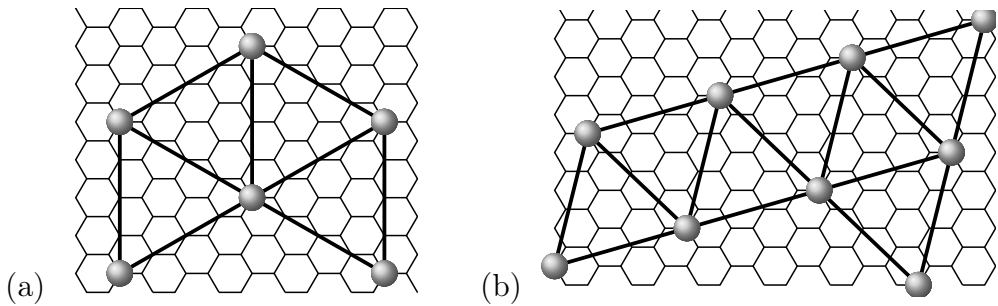 
When $D^2$ is not divisible by $3$, we pick the nearest L\"oschian number $(D^*)^2>D^2$
divisible by 3: the fact is that every MDA-sub-lattice is a
$D^*$-sub-lattice in $\bbH_2$, as triangles with area less than
that of the $D^*$-triangle do not generate $D$-admissible PGSs.
This defines a group of values $D^2$ called case HC, for which 
the above theory is repeated with $D$ replaced by $D^*$. See Theorems I (ii), 12 in \cite{MSS1}. Cf. Fig. 7.

\begin{figure}
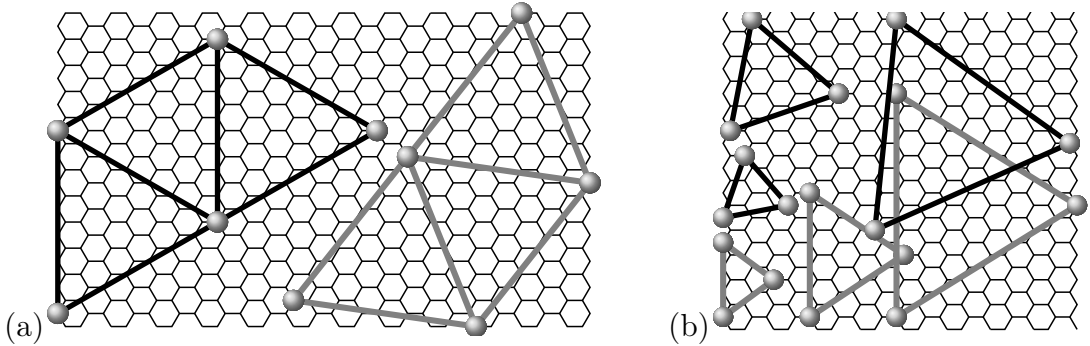

\begin{center} 
(a)
 \end{center}
\caption{(a) PGSs
on ${\mathbb H}_2$ for $D^2=147$ (Case HB). 
The `vertical' PGSs (black) are dominant. 
(b) PGSs on ${\mathbb H}_2$ for 
values $D^2$ from Class HC, with 
${D^*}^2=D^2+2$, and their associated ${D^*}$-triangles (black).
(i) For $D^2=19$,  ${D^*}^2=21$. 
(ii) For $D^2=61$, ${D^*}^2=63$. 
(iii) For $D^2=217$, ${D^*}^2=219$.
The gray ${\mathbb H}_2$-triangles give the minimal area when the side-lengths 
are $\geq D$ and the angles $\leq \pi/2$. However, they do not generate PGSs.
The PGSs are MDA-lattices constructed from the black ${D^*}$-triangles.}
\end{figure}  

\vskip .5cm
{\bf 3.5.} A different situation emerges on $\bbZ^2$, where
a $D$-triangle can never be inscribed. Nevertheless, every PGS-equivalence class is constructed from an MDA-sub-lattice.
The MDA-sub-lattices in $\bbZ^2$ are defined implicitly, via 
solutions of a discrete optimization problem
$$\begin{array}{c}\hbox{\bf minimize the area of a
${\bbZ}^2$-triangle \ $\triangle$}\\  \hbox{\bf with one vertex at the origin,}\\
\hbox{\bf with side-lengths \ $\ell_i\geq D$ and angles \
$\alpha_i\leq\pi/2$.}\end{array}\eqno (5)$$
The term $\bbZ^2$-triangle means a triangle with vertices
in $\bbZ^2$. Accordingly, $\sigma =S$ where $S/2$ is the
minimum achieved in (5). A minimizing triangle in (5) is referred
to as an {\it M-triangle}. Adjacent pairs of M-triangles form 
fundamental parallelograms of MDA-sub-lattices generating 
the PGSs. Cf. Fig. 8.

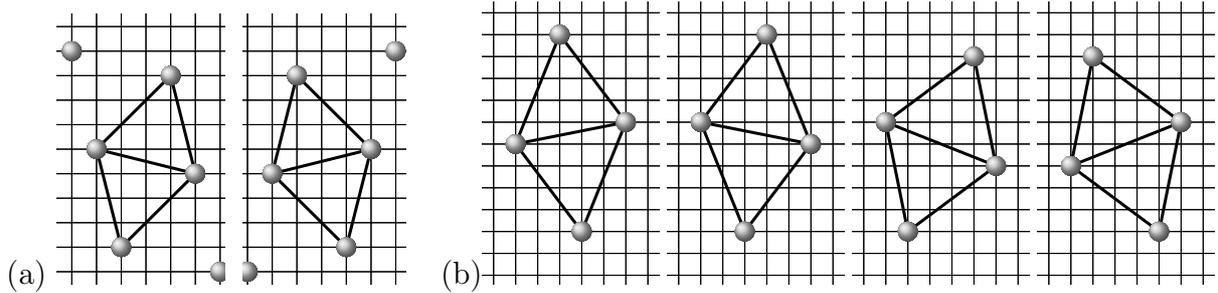
\begin{figure} \begin{center}
(a) \begin{tikzpicture}[scale=0.325]\label{2NSlidingD2=16}
\clip (-1.6, 1.5) rectangle (5.2, 12.5);


\path [draw=black, line width=0.4mm] (1,3)--(4,6)--(3,10)--(0,7)--(1,3);
\path [draw=black, line width=0.4mm] (4,6)--(0,7);

\draw[yscale=1, xslant=0] (-90,-60) grid (90, 60);
\draw[yscale=1, yslant=0] (-90,-60) grid (90, 60);
\foreach \pos in {(-1,11),(6,13),(0,7),(4,6),
(3,10),(1,3),(5,2)}
\shade[shading=ball, ball color=lightgray] \pos circle (.4);  
\end{tikzpicture}\;\;\begin{tikzpicture}[scale=0.325]
\clip (-0.2, 1.5) rectangle (6.4, 12.5);

\path [draw=black, line width=0.4mm] (1,6)--(4,3)--(5,7)--(2,10)--(1,6);
\path [draw=black, line width=0.4mm] (1,6)--(5,7);

\draw[yscale=1, xslant=0] (-90,-60) grid (90, 60);
\draw[yscale=1, yslant=0] (-90,-60) grid (90, 60);
\foreach \pos in {(2,10),(1,6),(5,7),(4,3),(0,2),(0,2),(6,11)}
\shade[shading=ball, ball color=lightgray] \pos circle (.4);  
\end{tikzpicture}\;\;\;\;(b)\begin{tikzpicture}[scale=0.29] \label{2NSlidingD2=25}
\clip (-1.5, -7.4) rectangle (6.5, 5.5);

\path [draw=black, line width=0.4mm] (0,-1)--(3,-5)--(5,0)--(2,4)--(0,-1);
\path [draw=black, line width=0.4mm] (0,-1)--(5,0);

\draw[yscale=1, xslant=0] (-90,-60) grid (90, 60);
\draw[yscale=1, yslant=0] (-90,-60) grid (90, 60);

\foreach \pos in {(0,-1),(3,-5),(2,4),(5,-0),(7,5)}
\shade[shading=ball, ball color=lightgray] \pos circle (.45);  
\end{tikzpicture}\;\begin{tikzpicture}[scale=0.29]
\clip (-1.5, -7.4) rectangle (6.5, 5.5);

\path [draw=black, line width=0.4mm] (0,0)--(3,4)--(5,-1)--(2,-5)--(0,0);
\path [draw=black, line width=0.4mm] (0,0)--(5,-1);

\draw[yscale=1, xslant=0] (-90,-60) grid (90, 60);
\draw[yscale=1, yslant=0] (-90,-60) grid (90, 60);

\foreach \pos in {(0,0),(3,4),(5,-1),(2,-5)}
\shade[shading=ball, ball color=lightgray] \pos circle (.45);  
\end{tikzpicture}\;\begin{tikzpicture}[scale=0.29]
\clip (-1.5, -7.4) rectangle (6.5, 5.5);

\path [draw=black, line width=0.4mm] (0,0)--(4,3)--(5,-2)--(1,-5)--(0,0);
\path [draw=black, line width=0.4mm] (0,0)--(5,-2);

\draw[yscale=1, xslant=0] (-90,-60) grid (90, 60);
\draw[yscale=1, yslant=0] (-90,-60) grid (90, 60);

\foreach \pos in {(0,0),(5,-2),(4,3),(1,-5)}
\shade[shading=ball, ball color=lightgray] \pos circle (.45);  
\end{tikzpicture}\;\begin{tikzpicture}[scale=0.29]
\clip (-1.5, -7.4) rectangle (6.5, 5.5);

\path [draw=black, line width=0.4mm] (0,-2)--(1,3)--(5,0)--(4,-5)--(0,-2);
\path [draw=black, line width=0.4mm] (0,-2)--(5,0);

\draw[yscale=1, xslant=0] (-90,-60) grid (90, 60);
\draw[yscale=1, yslant=0] (-90,-60) grid (90, 60);

\foreach \pos in {(0,-2),(1,3),(5,0),(4,-5),(0,-2)}
\shade[shading=ball, ball color=lightgray] \pos circle (.45);  
\end{tikzpicture}\end{center}
\caption{PGSs on ${\mathbb Z}^2$ for (a) $D^2=16$, $S=15$ 
and (b) $D^2=25$, $S=23$. In both cases, the M-triangles are ${\mathbb Z}^2$-congruent, 
and there is a single PGS-equivalence class. Consequently, $K=1$. For 
$D^2=16$ the M-triangles are isosceles, and there are 2 MDA-sub-lattices. For $D^2=25$ 
the M-triangles are non-isosceles, and there are 4 MDA-sub-lattices. 
Accordingly, $\sigma=15$, $m=2$ for $D^2=16$ and
$\sigma=23$, $m=4$ for $D^2=25$. The number of PGSs is 30 and 92, respectively.}
\end{figure}

Problem (5) always has a solution but the M-triangle  $\triangle$ may be non-unique. A delicate 
point is that there are different types of non-uniqueness of an M-triangle: (i) there 
are $N_0>1$ M-triangles, and they are $\bbR^2$- but not $\bbZ^2$-congruent (here 
$\bbZ^2$-congruent means that two M-triangles can be mapped into each other by a 
$\bbZ^2$-symmetry); (ii) there are $N_1>1$ M-triangles, and they are not $\bbR^2$-congruent; 
(iii) a mixture of (i) and (ii). Cf. Fig. 9. The value $K$ is the number of non-$\bbZ^2$-congruent 
M-triangles. Next: (a) $m_k=1$ if $D^2=2$, and (b) $m_k=2$ or $4$ when $D^2>2$ and the 
M-triangle $\triangle$ defining the PGS-equivalence class is isosceles or
not, respectively. See Theorems 1 (i), 2 and sections 3.1, 3.2 in \cite{MSS2}. 

Similarly to $\bbA_2$ and $\bbH_2$, the identification of PGSs on
$\bbZ^2$ is intrinsically connected with algebraic number
theory. 
It turns out that, for a given $D$, one can characterize the corresponding 
M-triangles via solutions to {\it norm equations} 
in ring $\bbZ [\sqrt[6]{-1}]$ (i.e., the ring of algebraic integers in the cyclotomic field generated by 
$\zeta_{12}=e^{\pi i/6}$). Such a connection helps to prove that uniqueness of an M-triangle and each of non-uniqueness types
(i)--(iii) occur for infinitely many values $D$, and the degrees of degeneracy 
$N_0$, $N_1$ can be arbitrarily large as $D\to\infty$. See Theorem 3 in \cite{MSS2}.

{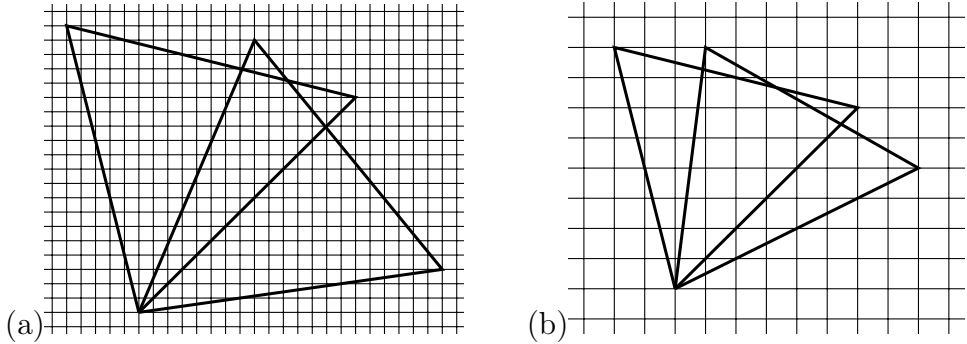
\begin{figure} \begin{center}
(a)\begin{tikzpicture}[scale=0.19]
\clip (-6.5, -1.5) rectangle (22.5, 21.5);


\draw[yscale=1, xscale=1] (-100,-100) grid (100, 100);

\path [draw=black, line width=0.4mm] (0,0) -- (15,15) -- (-5,20) -- (0,0);
\path [draw=black, line width=0.4mm] (0,0) -- (21,3) -- (8,19) -- (0,0);

\end{tikzpicture}\qquad (b)\begin{tikzpicture}[scale=0.4]
\clip (-3.5, -1.5) rectangle (9.5, 9.5);


\draw[yscale=1, xscale=1] (-100,-100) grid (100, 100);

\path [draw=black, line width=0.4mm] (0,0) -- (8,4) -- (1,8) -- (0,0);
\path [draw=black, line width=0.4mm] (0,0) -- (6,6) -- (-2,8) -- (0,0);
\end{tikzpicture} \end{center}
\caption{Non-uniqueness of M-triangles on ${\mathbb Z}^2$:
(a) for $D^2=425$, $S=375$ (${\mathbb R}^2$- but not ${\mathbb Z}^2$-congruent, $N_0=2$) 
and (b) for $D^2=65$, $S=60$ (${\mathbb R}^2$-non-congruent, $N_1=2$). Which PGS-class
generates EGDs is determined by dominance.}
\end{figure}} 
{\bf 3.6.}  It is possible to check that any non-periodic ground 
state on $\bbZ^2$ contains at least one infinite connected 
component of non-M-triangles and no finite ones. Moreover, the number of 
non-M-triangles in a $\bbZ^2$-square $\bbV (L)$ of side-length $L$ can only 
grow at most linearly with $L$; this means that in a non-periodic ground 
state, non-M-triangles form, effectively, a one-dimensional array. 

A similar pattern for non-periodic ground 
states emerges on $\bbA_2$ and $\bbH_2$. Let us repeat 
once more that non-periodic ground  states do not generate EGDs in dimension two \cite{DS}. Cf. discussions in section 3.1. 
 
\section{The Peierls bound}

{\bf 4.1. Contours.} Contours paly a central role in the P-S theory. Physically speaking, 
contours describe local perturbations of PGSs. They emerge when we (i) remove some
particles from a PGS $\vphi$ and (ii) add some new particles at `inserted' sites, maintaining 
$D$-admissibility.

A formal definition is as follows. First, we define a {\it template} as a parallelogram 
spanned by two non-collinear  vectors such that each MDA sub-lattice is invariant under shifts by 
both of these vectors. Typically we choose a parallelogram with a minimal area and a maximal acute angle.
Cf. Fig. 10.

{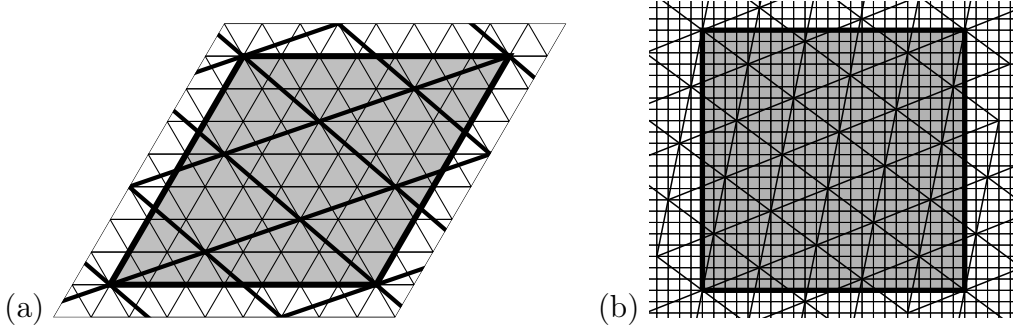
\begin{figure}\label{Fig10}\begin{center}
\captionsetup{width=0.8\textwidth, labelfont=bf}
\centering
(a)\;\begin{tikzpicture}[scale=0.5]

\filldraw [lightgray, yscale=sqrt(3/4), xslant=0.5] (0, 0) rectangle (7, 7);

\path [draw=black, line width=0.8mm, yscale=sqrt(3/4), xslant=0.5] (0, 0) rectangle (7, 7);

\clip[yscale=sqrt(3/4), xslant=0.5] (-1, -1) rectangle (8, 8);
\draw[yscale=sqrt(3/4), xslant=0.5] (-1,-1) grid (8, 8);
\draw[yscale=sqrt(3/4), xslant=-0.5] (-1,-1) grid (15, 8);


\begin{scope}
\def\n{5}
\def\aa{2}
\def\bb{1}
\inclinedgrid
\end{scope}\end{tikzpicture}\quad (b) \begin{tikzpicture}[scale=0.15]
\clip (-6.6, -15.4) rectangle (25.2, 12.5);
\filldraw[gray7] (-2,-13)--(-2,10)--(21,10)--(21,-13)--(-2,-13);

\path [draw=black, line width=0.7mm] (-2,-13)--(-2,10)--(21,10)
--(21,-13)--(-2,-13);

\path [draw=black, line width=0.2mm] (-9,-2)--(26,12); 
\path [draw=black, line width=0.2mm] (-5,-5)--(25,7);
\path [draw=black, line width=0.2mm] (-6,-10)--(24,2);
\path [draw=black, line width=0.2mm] (-7,-15)--(28,-1);
\path [draw=black, line width=0.2mm] (2,-16)--(27,-6);  
\path [draw=black, line width=0.2mm] (-8,3)--(22,15);   
\path [draw=black, line width=0.2mm] (-7,8)--(13,16);   
\path [draw=black, line width=0.2mm] (16,-15)--(21,-13)--(26,-11);   

\path [draw=black, line width=0.2mm] (-6,-10)--(-5,-5)--(-1,15); 
\path [draw=black, line width=0.2mm] (-3,-18)--(4,17);
\path [draw=black, line width=0.2mm] (2,-16)--(9,19);
\path [draw=black, line width=0.2mm] (6,-19)--(7,-14)--(8,-9)--(9,-4)
--(10,1)--(13,17); 
\path [draw=black, line width=0.2mm] (11,-17)--(12,-12)--(13,-7)
--(14,-2)--(15,3)--(17,13)--(18,18); 
\path [draw=black, line width=0.2mm] (16,-15)--(17,-10)--(18,-5)--(19,0)--(20,5)
--(21,10)--(22,15); 
\path [draw=black, line width=0.2mm] (20,-18)--(21,-13)--(22,-8)--(23,-3)
--(24,2); 

\path [draw=black, line width=0.2mm] (-6,-10)--(-2,-13)--(2,-16); 
\path [draw=black, line width=0.2mm] (-9,-2)--(-5,-5)--(-1,-8)--(3,-11)--(7,-14)--(11,-17);
\path [draw=black, line width=0.2mm] (-8,3)--(-4,0)--(0,-3)--(8,-9)--(12,-12)
--(16,-15); 
\path [draw=black, line width=0.2mm] (-7,8)--(-3,5)--(1,2)--(5,-1)--(9,-4)--(13,-7)
--(17,-10)--(21,-13)--(25,-16); 
\path [draw=black, line width=0.2mm] (-6,13)--(-2,10)--(2,7)--(6,4)--(10,1)
--(14,-2)--(18,-5)--(22,-8)--(26,-11); 
\path [draw=black, line width=0.2mm] (-1,15)--(3,12)--(7,9)--(11,6)--(15,3)--(23,-3)--(27,-6);
\path [draw=black, line width=0.2mm] (4,17)--(8,14)--(12,11)--(16,8)--(20,5)--(24,2);
\path [draw=black, line width=0.2mm] (13,16)--(17,13)--(21,10)--(25,7); 

\draw[yscale=1, xslant=0] (-90,-60) grid (90, 60);
\draw[yscale=1, yslant=0] (-90,-60) grid (90, 60);

\end{tikzpicture} \end{center}

\caption{Templates (gray) and fundamental  parallelograms on ${\mathbb A}_2$ for $D^2=7$
(a) and on ${\mathbb Z}^2$ for $D^2=25$ (b).}
\end{figure}} 
We say that a template $F$ is $\vphi$-regular in $\phi$ if, $\forall$ 
$x\in F$, we have $\phi (x)=\vphi (x)$. A template $F$ is called
$\vphi$-correct if $F$ and all 8 of its adjacent templates are
$\vphi$-regular. The {\it frustrated set} is formed by the union
of templates that are not $\vphi$-correct $\forall$ $\vphi\in
\cP (D)$.    

A contour $\Gam$ in a $D$-AC $\phi\in\cA (D)$ is defined as a
pair $(\rS,\phi\hskip -2pt\upharpoonright_\rS)$ where $\rS=\rSupp 
(\Gam )\subset\bbW$ is a connected component of the frustrated
set. We say that $\Gam$ is finite if the set $\rSupp (\Gam )$ is
finite. 

Let $\Gam$ be a finite contour in a $D$-AC $\phi\in\cA (D)$. The
complement $\bbW\setminus\rSupp (\Gam )$ has one infinite
connected component which we call the {\it exterior} of $\Gam$
and denote by Ext$\,(\Gam )$. In addition, set
$\bbW\setminus\rSupp (\Gam )$ may have finitely many 
{\it interior} connected components; they are denoted by Int$_j(\Gam )$, $j=1,\ldots ,J$, and we set
Int$\,(\Gam )=\operatornamewithlimits\cup\limits_{j=1}^J$Int$_j
(\Gam )$. Cf. Fig. 11. We say that $\Gam$ is a $\vphi$-contour in $\phi$ 
if every template $F\subset\rExt (\Gam )$ adjacent to 
$\rSupp (\Gam )$ is $\vphi$-correct in $\phi$. We say that 
$\Gam$ is an {\it external} contour in $\phi$ if $\rSupp (\Gam )$
does not lie in Int$\,(\Gam')$ for any other contour $\Gam'$. 


{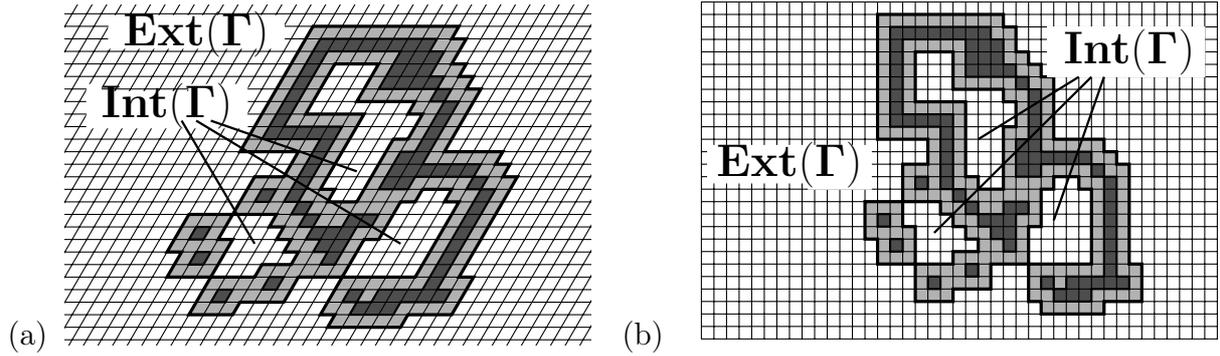
\begin{figure}\begin{center}
\captionsetup{width=0.8\textwidth, labelfont=bf}
\centering
(a)\;\;\begin{tikzpicture}[scale=0.192]
\clip (-2, 0.5) rectangle (34, 24);
\begin{scope}[yscale=sqrt(3/4), xslant=1/2]
\definecolor{gray3}{gray}{0.3} 
\definecolor{gray5}{gray}{0.7}
\path [fill=gray5, draw=black, very thick] (2, 6) -- (5, 6) -- (5, 9) -- (4, 9) -- (4, 11) -- (1, 11) -- (1, 8) -- (2, 8) -- cycle;
\path [fill=gray3, draw=black, thick] (3,7) -- (4, 7) -- (4, 8) -- (3, 8) -- cycle;
\path [fill=gray3, draw=black, thick] (2,9) -- (3, 9) -- (3, 10) -- (2, 10) -- cycle;
\path [fill=gray5, draw=black, very thick] (5,3) -- (8,3) -- (8,4) -- (11,4) -- (11,6) -- (14,6) -- (14,9) -- (15,9) --
(15,12) -- (16,12) -- (16,13) -- (18,13) -- (18,12) -- (19,12) -- (19,6) -- (14,6) -- (14,3) -- (15,3) -- (15,2) -- (20,2) --
(20,3) -- (23,3) -- (23,6) -- (22, 6) -- (22,14) -- (21,14) -- (21,15) -- (20,15) -- (20,16) -- (16,16) -- (16,21) -- (15,21) --
(15,22) -- (14,22) -- (14,23) -- (13,23) -- (13,25) -- (12,25) -- (12,26) -- (2,26) -- (2, 16) -- (6,16) -- (6,14) -- (4,14) --
(4,11) -- (7,11) -- (7,10) -- (8,10) -- (8,9) -- (9,9) -- (9,8) -- (10,8) -- (10,7) -- (8,7) -- (8,6) -- (5,6) -- cycle;
\path [fill=gray3, draw=black, thick] (6,4) -- (7,4) -- (7,5) -- (6,5) -- cycle;
\path [fill=gray3, draw=black, thick] (9,5) -- (10,5) -- (10,6) -- (9,6) -- cycle;
\path [fill=gray3, draw=black, thick] (5,12) -- (6,12) -- (6,13) -- (5,13) -- cycle;
\path [fill=gray3, draw=black, thick] (15,4) -- (16,4) -- (16,3) -- (19,3) -- (19,4) -- (22,4) -- (22,5) -- (21,5) -- (21,13)--
(20,13) -- (20,14) -- (19,14) -- (19,15) -- (15,15) -- (15,20) -- (14,20) -- (14,21) -- (13,21) -- (13,22) -- (12,22) -- (12,24)--
(11,24) -- (11,25) -- (3,25) -- (3,17) -- (7,17) -- (7,12) -- (8,12) -- (8,11) -- (9,11) -- (9,10) -- (10,10) -- (10,9) --
(11,9) -- (11,8) -- (12,8) --(12,7) -- (13,7) -- (13,10) -- (14,10) -- (14,11) -- (12,11) -- (12,10) -- (10,10) -- (10,11) --
(9,11) -- (9,12) -- (8,12) -- (8,13) -- (8,18) -- (4,18) -- (4,24) -- (9,24) -- (9,21) -- (13,21) -- (13,20) -- (14,20) --
(14,16) -- (13,16) -- (13,13) -- (15,13) -- (15,14) -- (19,14) -- (19,13) -- (20,13) -- (20,5) -- (17,5) -- (17,4) -- (16,4)--
(16,5) -- (15,5) -- cycle;
\path [fill=white, draw=black, very thick] (10,12) -- (12,12) -- (12,17) -- (13,17) -- (13,19) -- (12,19) -- (12,20) -- (8,20) --
(8,23) -- (5,23) -- (5,19) -- (9,19) -- (9,13) -- (10,13) -- cycle;
\draw (-16,0) grid (34,28);
\end{scope}
\path [fill=white, draw=white] (1,23.25)--(13,23.25)--(13,21)--(1,21)--cycle;
\def\boeta{{\mbox{\boldmath$\eta$}}}
\draw (7,22) node {\Large {${\bf{Ext}}({\mbox{\boldmath$\Gamma$}})$}};
\path [fill=white, draw=white] (-0.5,18.5)--(0.5,18.5)--(0.5,18.0)--(6.5,18.0)--(6.5,18.5)
--(10.5,18.5)--(10.5,15.5)--(-0.5,15.5)--cycle;
\draw (5,17) node {\Large {${\bf{Int}}({\mbox{\boldmath$\Gamma$}})$}};
\path [draw=black, line width=0.3mm] (8,16)--(18,12.5);
\path [draw=black, line width=0.3mm] (7,16)--(21,7.5);
\path [draw=black, line width=0.3mm] (6,16)--(11,7.5);
\end{tikzpicture}\quad  (b)\begin{tikzpicture}[scale=0.1655]
\clip (-15, -0.5) rectangle (29, 27);
\begin{scope}
\definecolor{gray3}{gray}{0.3}
\definecolor{gray5}{gray}{0.7}
\path [fill=gray5, draw=black, very thick] (2, 6) -- (5, 6) -- (5, 9) -- (4, 9) -- (4, 11) -- (1, 11) -- (1, 8) -- (2, 8) -- cycle;
\path [fill=gray3, draw=black, thick] (3,7) -- (4, 7) -- (4, 8) -- (3, 8) -- cycle;
\path [fill=gray3, draw=black, thick] (2,9) -- (3, 9) -- (3, 10) -- (2, 10) -- cycle;
\path [fill=gray5, draw=black, very thick] (5,3) -- (8,3) -- (8,4) -- (11,4) -- (11,6) -- (14,6) -- (14,9) -- (15,9) --
(15,12) -- (16,12) -- (16,13) -- (18,13) -- (18,12) -- (19,12) -- (19,6) -- (14,6) -- (14,3) -- (15,3) -- (15,2) -- (20,2) --
(20,3) -- (23,3) -- (23,6) -- (22, 6) -- (22,14) -- (21,14) -- (21,15) -- (20,15) -- (20,16) -- (16,16) -- (16,21) -- (15,21) --
(15,22) -- (14,22) -- (14,23) -- (13,23) -- (13,25) -- (12,25) -- (12,26) -- (2,26) -- (2, 16) -- (6,16) -- (6,14) -- (4,14) --
(4,11) -- (7,11) -- (7,10) -- (8,10) -- (8,9) -- (9,9) -- (9,8) -- (10,8) -- (10,7) -- (8,7) -- (8,6) -- (5,6) -- cycle;
\path [fill=gray3, draw=black, thick] (6,4) -- (7,4) -- (7,5) -- (6,5) -- cycle;
\path [fill=gray3, draw=black, thick] (9,5) -- (10,5) -- (10,6) -- (9,6) -- cycle;
\path [fill=gray3, draw=black, thick] (5,12) -- (6,12) -- (6,13) -- (5,13) -- cycle;
\path [fill=gray3, draw=black, thick] (15,4) -- (16,4) -- (16,3) -- (19,3) -- (19,4) -- (22,4) -- (22,5) -- (21,5) -- (21,13)--
(20,13) -- (20,14) -- (19,14) -- (19,15) -- (15,15) -- (15,20) -- (14,20) -- (14,21) -- (13,21) -- (13,22) -- (12,22) -- (12,24)--
(11,24) -- (11,25) -- (3,25) -- (3,17) -- (7,17) -- (7,12) -- (8,12) -- (8,11) -- (9,11) -- (9,10) -- (10,10) -- (10,9) --
(11,9) -- (11,8) -- (12,8) --(12,7) -- (13,7) -- (13,10) -- (14,10) -- (14,11) -- (12,11) -- (12,10) -- (10,10) -- (10,11) --
(9,11) -- (9,12) -- (8,12) -- (8,13) -- (8,18) -- (4,18) -- (4,24) -- (9,24) -- (9,21) -- (13,21) -- (13,20) -- (14,20) --
(14,16) -- (13,16) -- (13,13) -- (15,13) -- (15,14) -- (19,14) -- (19,13) -- (20,13) -- (20,5) -- (17,5) -- (17,4) -- (16,4)--
(16,5) -- (15,5) -- cycle;
\path [fill=white, draw=black, very thick] (10,12) -- (12,12) -- (12,17) -- (13,17) -- (13,19) -- (12,19) -- (12,20) -- (8,20) --
(8,23) -- (5,23) -- (5,19) -- (9,19) -- (9,13) -- (10,13) -- cycle;

\draw (-12,0) grid (30,28);
\end{scope}

\path [fill=white, draw=white] (-11.5,15.5)--(1.5,15.5)--(1.5,12)--(-3.5,12)--(-3.5,12.5)--(-11.5,12.5)--cycle;
\draw (-5,14) node {\Large{${\bf{Ext}}({\mbox{\boldmath$\Gamma$}})$}};
\path [fill=white, draw=white] (15.5,25)--(28,25)--(28,21)--(15.5,21)--cycle;
\draw (22,23) node {\Large{${\bf{Int}}({\mbox{\boldmath$\Gamma$}})$}};
\path [draw=black, line width=0.3mm] (18,21)--(10,16);
\path [draw=black, line width=0.3mm] (19,21)--(6.5,8.5);
\path [draw=black, line width=0.3mm] (20,21)--(16,9.5);
\end{tikzpicture} \end{center}
\caption{Contour supports on ${\mathbb A}_2/{\mathbb H}_2$ (a) and ${\mathbb Z}^2$ (b).  
Dark-gray color marks non-correct templates, light-gray templates are their neighbors. Note that 
both dark-gray and light-gray
templates are non-correct, and the difference between the two types is
 that
dark-gray are also non-regular, while the light-gray are regular.}
\end{figure}} 

An important point is that a contour can be considered without 
a reference
to the AC $\phi$: it is enough that we indicate (i) a $D$-AC
$\psi_\Gam$
over set $\rSupp (\Gam )$, (ii) an external phase $\vphi$ and 
the internal
phases $\vphi_j$, such that every template $F\subset\rExt 
(\Gam )$ adjacent
to $\rSupp (\Gam )$ is $\vphi$-correct and every template
$F\subset\rInt_j(\Gam )$ adjacent to $\rSupp (\Gam )$ 
is $\vphi_j$-correct, $j=1,\ldots ,J$. 

With the above definitions at hand, we write down the contour
representation for the partition function:
$$Z (\bbV_n\|\vphi )
=u^{\sharp (\vphi\upharpoonright_{\bbV_n})}
\sum_{\{\Gam_i\}\subset\bbV_n}\prod\limits_iw (\Gam_i).\eqno (6)$$
Here the summation goes over {\it compatible} contour collections
$\{\Gam_i\}$ with pair-wise disjoint $\rSupp (\Gam_i)\subset
\bbV_n$, while $w(\Gam )$ stands for the {\it statistical weight}
of contour $\Gam$:
$$w (\Gam )=u^{\sharp (\psi_\Gam )
-\sharp (\vphi\upharpoonright_\Gam )}. \eqno (7)$$  
Pictorially, compatibility means that if two 
of contours, $\Gam$ and $\Gam'$, from the collection
are not separated by a third contour $\ov\Gam$
then their external and/or internal phases are coordinated.
Formally, it requires two properties. (a) If (i) 
$\rSupp(\Gam')\subset\rInt_j
(\Gam )$ and (ii) there is no contour $\ov\Gam$ in $\{\Gam_i\}$
with $\rSupp(\Gam')\subset\rInt (\ov\Gam )$ and
$\rSupp(\ov\Gam)\subset\rInt_j (\Gam )$ then the internal phase 
$\vphi_j$ of $\Gam$ serves as the
external phase for $\Gam$. (b) If (i) $\rSupp(\Gam')\subset\rExt
(\Gam )$ and (ii) there is no contour $\ov\Gam$ in $\{\Gam_i\}$
with $\rSupp(\ov\Gam)\subset\rExt(\Gam )$ and
$\rSupp(\Gam')\subset\rInt(\ov\Gam )$ then $\Gam$ and $\Gam'$
have the same external phase. Note that all external contours in
a compatible collection $\{\Gam_i\}$ are $\vphi$-contours for 
the same $\vphi$. Here we refer to a $\vphi$-contour when the external 
phase of the contour coincides with $\vphi$, i.e., all corresponding
external light-gray squares in Fig. 11 are $\vphi$-correct.

The condition upon $\bbV_n$ in (6) is that $\bbV_n$ is a finite
union of templates.

We then can think of a Gibbs distribution
$\mu_{\bbV_n}(\,\cdot\,\|\vphi )$ as a probability measure on
compatible contour collections $\{\Gam_i\}$ in $\bbV_n$. 
\vskip 0.5cm

{\bf 4.2. The Peierls constant.} The next step is to establish 
a bound 
$$\sharp (\psi_\Gam )-\sharp (\vphi\hskip -3pt\upharpoonright_\Gam )\leq 
-p\|\rSupp (\Gam )\|.\eqno (8)$$
Here  $\|\rSupp (\Gam )\|$ stands for the number of templates 
in $\rSupp (\Gam )$ and $p=p(D,\bbW )>0$ is a {\it Peierls
constant} per a template. Then the statistical weight $w (\Gam )$
will obey $w (\Gam )\leq u^{-p\|\rSupp (\Gam )\|}$, and a
standard Peierls argument will lead to properties ({\bf P1-5})
below.

By the definition of a contour, every template $F\subset\rSupp 
(\Gam )$ contains some sort of a defect. A trivial defect is 
when at least one particle can be added to configuration  
$\psi_\Gam\hskip -2pt\upharpoonright_F$; in this case we get at least a 
factor 1 in place of $p$ in inequality (8). 

For a saturated defective template $F\subset\rSupp (\Gam )$, 
the situation is more complex. Here we use the {\it Delaunay 
triangulation} for $\psi_\Gam\hskip -4pt\upharpoonright_F$; the fact is that
for each of the triangles the area is $\geq\sigma /2$, and for 
at least one triangle the area is $\geq (1+\sigma )/2$. (Note that the
lattice triangle has a half-integer area.) The number of such
triangles is $O\left(\|\rSupp (\Gam )\|\right)$. On the other
hand, the number of particles inside $\rSupp (\Gam )$ is twice the 
number of triangles. This ultimately leads to (8).

The constant $p (D)$ can actually be lover-bounded by 
$c/D^2$.

\section{Dominance, extreme Gibbs distributions}

{\bf 5.1.} If for a given $D$, the number $\sharp \cE (D, \mathbb{W})$ of PGS-equivalence
classes
equals 1, then for $u$ large enough the number of EGDs  equals
the number $\sharp \cP (D, \mathbb{W})$ of PGSs. When the number of PGS-equivalence
classes $K$ is greater than $1$,
i.e., there are multiple PGS-equivalence classes, not all
PGS-classes generate EGDs. In a sense, it is an expected behavior since an absence of explicit 
symmetry between the representatives of different PGS-equivalence classes results in different 
sets of their perturbations (contours). The PGSs generating EGDs are referred to as stable/dominant ones. The dominant
classes are those which minimize a `truncated free energy' for the 
ensemble of small contours; cf. \cite{Z1}. Observe that typically in the P-S theory 
the statistical weight of a contour decreases exponentially with the size of the contour.
Accordingly, the contours can be listed in the decreasing order of their statistical weights;
geometrically smaller contours appear earlier in the list. A truncated free energy emerges 
when we discard the contours whose statistical weight is lower than a given threshold value. 

We conjecture that for the H-C model the dominant PGS-equivalence class is always unique, since otherwise there would be too many additional hidden symmetries. 

\vskip .5cm

{\bf 5.2.} As an outcome of the P-S theory for the H-C model we have the following properties.  

\smallskip
({\bf P1}) Each EGD $\bmu\in\cE (D,\mathbb{W})$ is generated 
by a PGS. That is, each EGD is of the form $\bmu_\vphi$ for 
some $\vphi\in\cP (D,\mathbb{W})$. If PGSs $\vphi_i$ generate
EGDs $\bmu_{\vphi_i}$, $i=1,2$, and $\vphi_1\neq\vphi_2$ then
$\bmu_{\vphi_1}\perp\bmu_{\vphi_2}$. The EGDs inherit symmetries
from the PGSs. 

({\bf P2}) Consequently, EGD-generation is a class property: if
a PGS $\vphi\in\cP (D,\mathbb{W})$ generates an EGD $\bmu_\vphi$
then every PGS ${\vphi'}$ from the same equivalence 
class generates an EGD $\bmu_{\vphi'}$. Such a class is referred
to as dominant. Obviously, if a PGS-equivalence class is unique, it is dominant.

({\bf P3}) Each EGD $\bmu_\vphi$ exhibits the following 
properties. For $\bmu_\vphi$-almost all $\phi\in\cA$:
(i) all contours $\Gam$ in $\phi$ are finite, (ii) for 
any site $x\in\bbW$ there exist only finitely many contours
$\Gam$ (possibly none) such that 
$x$ lies in the interior of \, $\Gam$, (iii) there are countably 
many disjoint $D$-connected sets of $\vphi$-correct templates, 
one of which is infinite while all remaining $D$-connected
sets are finite, (iv) for every $\vphi'\in\cP
(D,\mathbb{W})\setminus\{\vphi\}$,
there are countably many $D$-connected sets of $\vphi'$-correct
templates all of which are finite. 

({\bf P4}) EGD $\bmu_\vphi$ admits a polymer expansion and 
has an exponential decay of correlations.

({\bf P5}) As $u\to\infty$, $\bmu_\vphi$
converges weakly to a measure sitting on a single $D$-AC $\vphi$.

\smallskip
Property ({\bf P3}) establishes a percolation picture for EGD
$\bmu_\vphi$: there is a `sea' of $\vphi$-correct templates
with `islands' of non-$\vphi$-correct sites inside which there 
may be `lakes' of templates of different correctness,
etc. In such a picture, contours mark `coastal/shallow-water
strips'. \vskip 0.5cm

{\bf 5.3.} Specification of dominant PGS-classes for a general value $D$ remains an open question; in \cite{MSS1}, Theorems 4, 5, 6, 10, we 
present a complete answer for values $D^2=$ 49, 147, 169 on $\bbA_2$ or $\bbH_2$ by classifying 
small contours. We expect that a similar approach could be used for other examples.

\begin{figure}[H]
\begin{center}
\captionsetup{width=0.8\textwidth, labelfont=bf}
\centering
(a) \begin{tikzpicture}[scale=0.27]
\clip (-1.2,-3.2*\rr) rectangle (18,20.5*\rr);

\path [draw=black,line width=0.3mm]
(0,0) arc (128.21:68.21:14*\rr) arc (248.21:188.21:14*\rr) arc (368.21:308.21:14*\rr)
(12, 2*\rr) arc (188.21:128.21:14*\rr) arc (308.21:248.21:14*\rr) arc (68.21:8.21:14*\rr);

\path [fill=lightgray, draw=black] (12, 2*\rr) arc (248.21:188.21:14*\rr) arc (68.21:8.21:14*\rr);
\path [fill=lightgray, draw=black] (0,0) arc (128.21:68.21:14*\rr) (12, 2*\rr)--(0,0);
\path [fill=lightgray, draw=black] (12,2*\rr) arc (188.21:128.21:14*\rr) (16.5, 15*\rr)--(12,2*\rr);
\path [fill=lightgray, draw=black]  (16.5, 15*\rr) arc (308.21:248.21:14*\rr) (4.5,13*\rr) 
--(16.5, 15*\rr);
\path [fill=lightgray, draw=black] (4.5,13*\rr) arc (368.21:308.21:14*\rr);

\path [draw=black, line width=0.4mm]
(0,0) -- (4.5,13*\rr) -- (16.5, 15*\rr) -- (12, 2*\rr) -- cycle
(12, 2*\rr) -- (4.5,13*\rr) -- cycle;

\draw[yscale=sqrt(3/4), xslant=0.5] (-90,-60) grid (90, 60);
\draw[yscale=sqrt(3/4), xslant=-0.5] (-90,-60) grid (90, 60);

\foreach \pos in {(0,0), (4.5,13*\rr),(12, 2*\rr),(16.5, 15*\rr)}
\shade[shading=ball, ball color=white] \pos circle (.5);

\foreach \pos in {(2, 2*\rr),(3, 2*\rr),   
(2.5, 3*\rr),(3.5, 3*\rr),(4.5, 3*\rr),(5.5, 3*\rr),(6.5, 3*\rr),(7.5, 3*\rr),(8.5, 3*\rr),(9.5, 3*\rr), 
(3,4*\rr),(4,4*\rr),(5,4*\rr),(6,4*\rr),(7,4*\rr),(8,4*\rr),(9,4*\rr), 
(3.5,5*\rr),(4.5,5*\rr),(5.5,5*\rr),(6.5,5*\rr),(7.5,5*\rr), 
(4,6*\rr),(5,6*\rr),(6,6*\rr),(7,6*\rr), 
(4.5,7*\rr),(5.5,7*\rr),(6.5,7*\rr), 
(5,8*\rr),(6,8*\rr),  
(4.5,9*\rr),(5.5,9*\rr),(5,10*\rr),  
(11.5,5*\rr),(11,6*\rr),  (12,6*\rr),  
(10.5,7*\rr),(11.5,7*\rr), 
(10,8*\rr),(11,8*\rr),(12,8*\rr),  
(9.5,9*\rr),(10.5,9*\rr),(11.5,9*\rr),(12.5,9*\rr),  
(9,10*\rr),(10,10*\rr),(11,10*\rr),(12,10*\rr),(13,10*\rr),  
(7.5,11*\rr),(8.5,11*\rr),(9.5,11*\rr),(10.5,11*\rr),(11.5,11*\rr),(12.5,11*\rr),(13.5,11*\rr), 
(7,12*\rr),(8,12*\rr),(9,12*\rr),(10,12*\rr),(11,12*\rr),(12,12*\rr),(13,12*\rr),(14,12*\rr),  
(13.5, 13*\rr),(14.5, 13*\rr)}  
\filldraw[darkgray] \pos circle (.3);
\end{tikzpicture}\qquad (b) \begin{tikzpicture}[scale=0.27]
\clip (-4.2, -1.2*\rr) rectangle (15, 22.5*\rr);

\path [draw=black,line width=0.3mm]
(0, 0) arc (-30:30:14*\rr) arc (210:270:14*\rr) arc (90:150:14*\rr)
(10.5, 7*\rr) arc (30:90:14*\rr) arc (270:330:14*\rr) arc (150:210:14*\rr);

\path [fill=lightgray, draw=black] (0, 14*\rr) arc (210:270:14*\rr) arc (30:90:14*\rr);
\path [fill=lightgray, draw=black] (0, 0) arc (-30:30:14*\rr) (0, 14*\rr)--(0, 0);
\path [fill=lightgray, draw=black] (10.5, 7*\rr)  arc (90:150:14*\rr) (0, 0)--(10.5, 7*\rr);
\path [fill=lightgray, draw=black] (10.5, 21*\rr) arc (150:210:14*\rr) (10.5, 7*\rr)
--(10.5, 21*\rr);
\path [fill=lightgray, draw=black] (0, 14*\rr) arc (270:330:14*\rr)
(10.5, 21*\rr)--(0, 14*\rr);
 
\draw [line width=0.4mm] (0, 0) -- (0, 14*\rr) -- (10.5, 21*\rr) -- (10.5, 7*\rr) -- cycle
(0, 14*\rr) -- (10.5, 7*\rr) -- cycle;

\draw[yscale=sqrt(3/4), xslant=0.5] (-19,-2) grid (90, 60);
\draw[yscale=sqrt(3/4), xslant=-0.5] (-5,-2) grid (90, 60);

\foreach \pos in {(0, 0), (10.5, 7*\rr),(0, 14*\rr),(10.5, 21*\rr)}
 \shade[shading=ball, ball color=white] \pos circle (.5);

\foreach \pos in {(0.5,1*\rr),(1,2*\rr),(1.5,3*\rr), 
(2,4*\rr),(3,4*\rr),  
(1.5,5*\rr),(2.5,5*\rr),(3.5,5*\rr),  
(2,6*\rr),(3,6*\rr),(4,6*\rr),(5,6*\rr),(6,6*\rr),  
(2.5,7*\rr),(3.5,7*\rr),(4.5,7*\rr),(5.5,7*\rr),(6.5,7*\rr), 
(7.5,7*\rr),(8.5,7*\rr),(9.5,7*\rr),  
(2,8*\rr),(3,8*\rr),(4,8*\rr),(5,8*\rr),(6,8*\rr), 
(1.5,9*\rr),(2.5,9*\rr),(3.5,9*\rr),  
(2,10*\rr),(3,10*\rr),  
(1.5,11*\rr),(1,12*\rr),(0.5,13*\rr),  
(10,8*\rr),(9.5,9*\rr),(9,10*\rr),  
(7.5,11*\rr),(8.5,11*\rr),  
(7,12*\rr),(8,12*\rr),(9,12*\rr),  
(4.5,13*\rr),(5.5,13*\rr),(6.5,13*\rr),(7.5,13*\rr),(8.5,13*\rr), 
(1,14*\rr),(2,14*\rr),(3,14*\rr),(4,14*\rr),(5,14*\rr),(6,14*\rr),(7,14*\rr),(8,14*\rr), 
(4.5,15*\rr),(5.5,15*\rr),(6.5,15*\rr),(7.5,15*\rr),(8.5,15*\rr),  
(7,16*\rr),(8,16*\rr),(9,16*\rr),  
(7.5,17*\rr),(8.5,17*\rr), 
(9,18*\rr),(9.5,19*\rr),(10,20*\rr)} 
\filldraw[darkgray] \pos circle (.3);
\end{tikzpicture}\vskip .3cm

(c) \begin{tikzpicture}[scale=0.24]
\clip (-1.2, -2.2*\rr) rectangle (26.2, 29.2*\rr);
\draw[yscale=sqrt(3/4), xslant=0.5] (-90,-60) grid (90, 60);
\draw[yscale=sqrt(3/4), xslant=-0.5] (-90,-60) grid (90, 60);

\path [draw=black, line width=0.5mm, color=black] (11.5,5*\rr)--(9.5,21*\rr);
\foreach \pos in {(11.5,5*\rr),(9.5,21*\rr)}
\shade[shading=ball, ball color=black] \pos circle (.5);

\path [draw=black, line width=0.5mm, color=black] (2,2*\rr)--(11,12*\rr)--(21.5,5*\rr);
\path [draw=black, line width=0.5mm, color=black] (9.5,23*\rr)--(12.5,9*\rr)--(2,2*\rr);

\foreach \pos in {(2,2*\rr),(21.5,5*\rr),(11,12*\rr)}
 \shade[shading=ball, ball color=black] \pos circle (.5);
\foreach \pos in {(9.5,23*\rr),(12.5,9*\rr)}
 \shade[shading=ball, ball color=black] \pos circle (.5);

 \path [draw=black, line width=0.3mm] (0,0) -- (9, 26*\rr) -- (24, 4*\rr) -- cycle
(12, 2*\rr) -- (4.5,13*\rr) -- (16.5, 15*\rr) -- cycle;

\foreach \pos in {(0,0),(4.5,13*\rr),(9, 26*\rr),(12, 2*\rr),(24, 4*\rr),(16.5, 15*\rr)}
 \shade[shading=ball, ball color=white] \pos circle (.6);

\end{tikzpicture}\qquad (d) \begin{tikzpicture}[scale=0.24]
\clip (-1.2, -1.2*\rr) rectangle (23.2, 30.2*\rr);
\draw[yscale=sqrt(3/4), xslant=0.5] (-19,-2) grid (90, 60);
\draw[yscale=sqrt(3/4), xslant=-0.5] (-1,-2) grid (90, 60);

\path [draw=black, line width=0.5mm, color=black] (3.5,5*\rr)--(10,20*\rr);
\foreach \pos in {(3.5,5*\rr),(10,20*\rr)}
\shade[shading=ball, ball color=black] \pos circle (.5);

\draw [line width=0.5mm, color=black] (0.5, 1*\rr)--(7.5,15*\rr)--(0.5, 27*\rr);
\foreach \pos in {(0.5, 1*\rr),(7.5,15*\rr),(0.5, 27*\rr),}
 \shade[shading=ball, ball color=black] \pos circle (.5); 

\draw [line width=0.5mm, color=black] (0.5,1*\rr)--(6,14*\rr)--(0.5,27*\rr);
\draw [line width=0.5mm, color=black] (6,14*\rr)--(19,14*\rr);
 \foreach \pos in {(6, 14*\rr),(19, 14*\rr)}
 \shade[shading=ball, ball color=black] \pos circle (.5);
\foreach \pos in {(0.5, 1*\rr),(0.5, 27*\rr)}
 \shade[shading=ball, ball color=black] \pos circle (.5);
 
\draw [line width=0.3mm] (0,0) -- (0, 28*\rr) -- (21, 14*\rr) -- cycle
(0, 14*\rr) -- (10.5, 7*\rr) -- (10.5, 21*\rr) -- cycle;

\foreach \pos in {(0, 0), (10.5, 7*\rr),(21, 14*\rr),(0, 14*\rr),(0, 28*\rr),(10.5, 21*\rr)}
 \shade[shading=ball, ball color=white] \pos circle (.6);\end{tikzpicture}
\end{center}
\caption{Dominance on ${\mathbb A}_2$, for $D^2=147$. This figure shows admissible 
$u^{-2}$-insertions for $D^2=147$ on $\mathbb{A}_2$. Here we have two PGS-equivalence 
classes, one `inclined' (a, c) and one `vertical' (b, d). Single admissible $u^{-2}$-insertions 
are marked by dark dots (a, b); gray lenses indicate areas from where a single insertion 
repels four sites in a PGS and hence yields a $u^{-2}$-insertion. The number of single 
$u^{-2}$-insertions is 68 per a $D$-rhombus for both PGS-classes. Double and triple 
$u^{-2}$-insertions are shown on frames (c, d) and a quadruple insertion on frame (d).
For the inclined class there is no admissible quadruple $u^{-2}$-insertion, and for the 
neither class there is an admissible $n$-particle $u^{-2}$-insertion with $n\geq 5$. The 
count of all $u^{-2}$-insertions shows that the vertical  PGS-class is dominant (an 
argument presented in \cite{MSS1}, sections 6, 7, shows that contributions of order $u^{-3}$ and higher 
are irrelevant in the case under consideration).}
\end{figure}
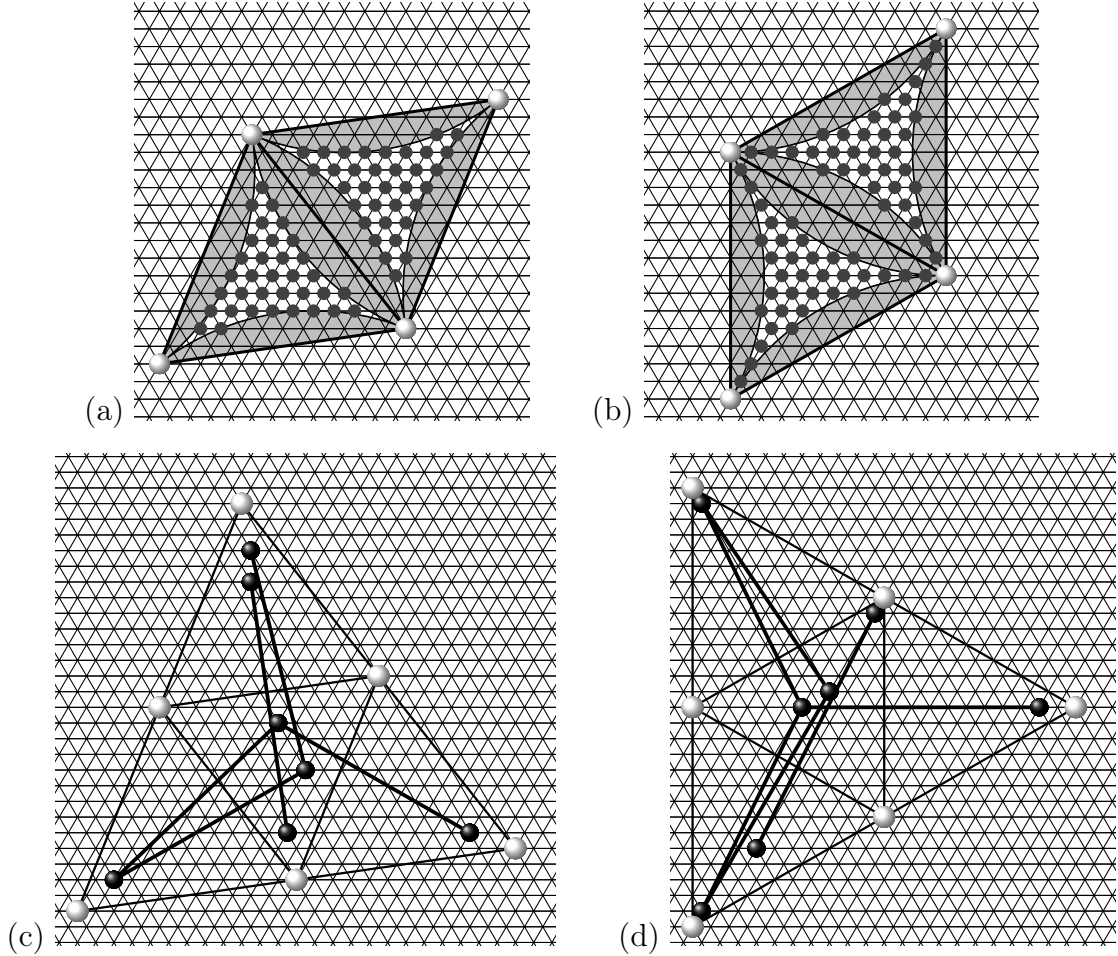 

The simplest small contour occurs when we merely remove a particle from a PGS $\vphi$. Clearly, all PGSs 
are `equal in rights' relative to this perturbation type, as they are  maximally-dense. The same is true about 
removing pairs of particles, etc. Then there is a possibility of inserting one and removing three particles, at 
the vertices of the Delaunay triangle of $\vphi$; 
these are single insertions. Note that inserting a single particle implies removing 
at least three. Continuing, we can think of double insertions where pairs of inserted particles remove four 
particles at the vertices of covering parallelogram in $\vphi$, and so on,  maintaining the difference 2 between 
the number of removed and inserted particles (and, of course, $D$-admissibility). This yields a finite collection 
of admissible $u^{-2}$-{\it insertions} where we add $n$ and remove $n+2$ particles.

In the above-mentioned examples,  
we ({\bf a}) enumerate the $u^{-2}$-insertions, and ({\bf b}) verify that higher-order 
insertions (beginning with $u^{-3}$) can be discarded when fugacity $u$ is large enough. Task ({\bf b}) 
is achieved via standard polymer expansion techniques. Accordingly, the free energy 
of the model under consideration is represented as an absolutely convergent polymer series, where the 
polymers are connected unions of contours. The amount of such objects of size $n$ is upper-bounded 
by $C^n$, where $C$ is an absolute constant and $n$ is the number of the lattice sites in a polymer. On the 
other hand, the absolute value of the statistical weight of a polymer is upper-bounded by $u^{-pn}$, where $p$ 
is the Peierls constant from (8). Therefore, the polymer expansion converges for large $u$ as soon as $u^{-p}C<1$. 
Moreover, for all $u>2C^{1/p}$ there exists $n_0$ such that the tail of the polymer 
series (with $n\geq n_0$) does not exceed $u^{-3}\ll u^{-2}$ in the absolute value. The remaining initial 
part of the series does not contain more than $C^{2n_0}$ terms. For $u> C^{4n_0}$ the total contribution of terms 
with the statistical weight less than or equal to $u^{-3}$ does not exceed $u^{-3}C^{2n_0}< u^{-2.5}\ll 
u^{-2}$.

This way the problem is reduced to task ({\bf a}): enumeration of the $u^{-2}$-insertions. 
The most involved part (requiring a computer-assisted proof) is to check that the maximal number 
of added particles in admissible $u^{-2}$-insertions is $n=4$: here we remove $6$ occupied sites 
from the boundary of a $2D$-triangle in a given PGS; cf. Fig. 12 (d) for $D^2=147$. Generally, a 
Peierls bound implies that all $u^{-2}$-insertions have bounded size depending on the corresponding 
constant $p$. The problem is that $p$ is relatively small, and the corresponding 
bound is impractically large. The fact that 
there is no insertion of $n\geq 5$ particles implying the removal of at most $n+2$ particles 
requires an additional argument. See \cite{MSS1}, Lemmas 7.1--7.3.

The above approach allows us to identify the dominant PGS-classes for $D^2=49, 147, 169$ on $\bbA_2$ 
and $\bbH_2$.

\section{Conclusions}

In this work we construct the phase diagram for the hard-core model on $\bbA_2$, $\bbH_2$ 
and $\bbZ^2$ in a high-density/large-fugacity regime for all non-sliding values of the 
exclusion distance $D$ or, equivalently, for any label $k$ of excluded neighbors. These results are rigorously established in the thermodynamic limit. The H-C model 
in this regime admits a  straightforward application of the P-S theory, but the challenge here 
lies in determining the periodic ground states and verifying a suitable Peierls bound. A zest of 
the work is that the PGSs have been identified by means of algebraic number theory. Since the 
PGSs coincide with periodic disk-packings of maximal density, we also obtain the description of 
the latter, which yields a solution to the disk-packing problem on $\bbA_2$, $\bbH_2$ and 
$\bbZ^2$ for all disk diameters $D>0$. 

In particular, the complete structure of the large fugacity-phase diagram has been established when 
there is a single PGS-equivalence class. Also, it has been proven that there are only finitely many 
sliding values of $D$, and all of them have been identified.

The work also highlights open problems beyond the P-S theory. (1) How to identify 
the dominant class(es) among multiple PGS-equivalence classes? (2) What is the phase diagram 
as a function of $u>0$ for a given value of $D$? 
(3) What is the dependence of the critical value(s) of $u$ upon $D$? If point $u^0=u^0(D)$ separates the 
uniqueness and non-uniqueness domains, and $u^1(D)$ is the lower threshold for the large-fugacity regime, 
then do we have $u^0(D)=u^1(D)$ for some/all/no $D$? In other words, is there one or several
phase transitions for a given $D$ and what is the type of the transition(s)?  (3) Could our results 
provide hints towards understanding the model in $\bbR^2$? The challenge here is 
that our technique is applicable for $u>u^0(D)$, where $u^0(D)$ increases to $\infty$ as $D\nearrow \infty$.



A natural next step is to consider the HC model on $\bbZ^3$. Here the situation 
is more complex, mainly due to its close relationship with the Kepler 
conjecture. Nevertheless, similar P-S theory based results have been rigorously 
obtained in \cite{MSS3} for certain special values of $D$: $D^2=2, 3, 4, 5, 6, 8, 9, 
10, 12, 2\ell^2$ where $\ell\in \bbZ$. Also see \cite{JTMSR}, \cite{VMDDR}.

\vskip .5cm

{\bf Acknowledgement.} {IS and YS thank the Math Department, Penn State 
University, for support during this work. YS thanks St John's College, Cambridge, for support.
The authors express their gratitude to the referees for numerous useful remarks and suggestions.

\end{document}